\documentclass[authoryear,review,preprint,12pt]{elsarticle}
\usepackage[english,french]{babel}    
\usepackage[T1]{fontenc} 
\usepackage[cyr]{aeguill}
\usepackage{epsfig}
\usepackage{graphicx} 
\usepackage{epstopdf}
\usepackage{lscape} 
\usepackage[utf8]{inputenc}
\usepackage{multirow}
\usepackage{amsmath,amssymb,amsthm,mathrsfs,amsfonts,dsfont}
\usepackage[nottoc, notlof, notlot]{tocbibind} 
\usepackage{a4wide}
\usepackage{url}
\usepackage{color}
\usepackage{natbib}

\usepackage{lipsum}
\usepackage{multicol}
\usepackage{nopageno}
\usepackage{amsmath}
\usepackage{array}
\usepackage{amsthm}
\usepackage{numprint}
\usepackage{float}
\usepackage{amsfonts}
\usepackage{lineno}
\usepackage{tikz}
\usetikzlibrary{matrix, shapes, arrows, positioning}

\tikzset{>=latex}

\tikzstyle{decision} = [diamond, draw, fill=green!20, 
    text width=6em,minimum height=2em, text centered, node distance=.12\linewidth,]
\tikzstyle{block} = [rectangle, draw, fill=blue!20, 
    text width=16em, text centered, rounded corners, minimum height=2em,node distance=.12\linewidth,]
\tikzstyle{block2} = [rectangle, draw, fill=red!20, 
    text width=10em, text centered, rounded corners, minimum height=2em,node distance=.12\linewidth,]
\tikzstyle{block3} = [rectangle, draw=red,fill=white!20, dashed,
    text width=10em, text centered, rounded corners, minimum height=2em,node distance=.12\linewidth,]

\tikzstyle{line} = [draw, -latex',dotted]
\tikzstyle{cloud} = [draw, ellipse,fill=red!20, node distance=0.1\linewidth,
    minimum height=2em]

\usepackage{numcompress}
\usepackage{fullpage}
\usepackage{empheq}
\usepackage{amssymb}
\usepackage{remreset}
\makeatletter \@removefromreset{footnote}{chapter} \makeatother
\usepackage {geometry}
\usepackage{subfigure}
\usepackage{listingsutf8}

\usepackage{wrapfig}
\begin{document}
\selectlanguage{english}

\begin{frontmatter}
\title{Theory for planetary exospheres: II. Radiation pressure effect on exospheric density profiles}
\author[ups,irap]{A.~Beth\corref{cor1}}
\ead{arnaud.beth@gmail.com}
\author[ups,irap]{P.~Garnier\corref{cor2}}
\ead{pgarnier@irap.omp.eu}
\author[ups,irap]{D.~Toublanc}
\author[ups,irap]{I.~Dandouras}
\author[ups,irap]{C.~Mazelle}
\address[ups]{Université de Toulouse; UPS-OMP; IRAP; Toulouse, France}
\address[irap]{CNRS; IRAP; 9 Av. colonel Roche, BP 44346, F-31028 Toulouse cedex 4, France}

\cortext[cor1]{Principal corresponding author}
\cortext[cor2]{Corresponding author}
\begin{abstract}
The planetary exospheres are poorly known in their outer parts, since the neutral densities are low compared with the instruments detection capabilities. The exospheric models are thus often the main source of information at such high altitudes. We present a new way to take into account analytically the additional effect of the radiation pressure on planetary exospheres. In a series of papers, we present with an Hamiltonian approach the effect of the radiation pressure on dynamical trajectories, density profiles and escaping thermal flux. Our work is a generalization of the study by \citet{Bishop1989}. In this second part of our work, we present here the density profiles of atomic Hydrogen in planetary exospheres subject to the radiation pressure. We first provide the altitude profiles of ballistic particles (the dominant exospheric population in most cases), which exhibit strong asymmetries that explain the known geotail phenomenon at Earth. The radiation pressure strongly enhances the densities compared with the pure gravity case (i.e. the Chamberlain profiles), in particular at noon and midnight. We finally show the existence of an exopause that appears naturally as the external limit for bounded particles, above which all particles are escaping. 
\end{abstract}

\end{frontmatter}
\section{Introduction}
The exosphere is the upper layer of any planetary atmosphere: it is a quasi-collisionless medium where the particle trajectories are more dominated by gravity than by collisions. 
 Above the exobase, the lower limit of the exosphere, the Knudsen number \citep{Ferziger1972} becomes large, collisions become scarce, the distribution function cannot be considered as maxwellian anymore and, gradually, the trajectories of particles are essentially determined by the gravitation and radiation pressure by the Sun. The trajectories of particles, subject to the gravitational force, are completely solved with the equations of motion, but it is not the case with the radiation pressure \citep{Bishop1989}.

To describe correctly the exospheric population, we distinguish three Types of particles: escaping, ballistic and satellite (\citet{Chamberlain1963}, \citet{Banks1973}).

\begin{itemize}
\item the escaping particles come from the exobase and have a positive mechanical energy: they can escape from the gravitational influence of the planet with a velocity larger than the escape velocity. These particles are responsible for the Jeans' escape \citep{Jeans1916}.
\item the ballistic particles also come from the exobase but with a negative mechanical energy, they are gravitationally bound to the planet. They reach a maximum altitude and fall down on the exobase if they do not undergo collisions.
\item the satellite particles never cross the exobase. They also have a negative mechanical energy but their periapsis is above the exobase: they orbit along an entire ellipse around the planet without crossing the exobase. The satellite particles result from ballistic particles undergoing few collisions mainly near the exobase. Thus, they do not exist in a collisionless model of the exosphere.
\end{itemize}

By definition, their trajectories are conics in the pure gravity case. \citet{Chamberlain1963} proposed an analytical approach to estimate the density of each population via Liouville's theorem which states that the distribution function remains constant along a dynamical trajectory. A maxwellian distribution function is assumed at the exobase and propagated to the upper layers via Liouville's theorem. The density for each population is then derived as the product between the barometric law and a partition function $\zeta$.
\begin{align}
n(r)&={}n_{bar}\zeta(\lambda)\label{Chamberlain}\\
\nonumber&={}n(r_{exo}) e^{\lambda-\lambda_{exo}}(\zeta_{bal}+\zeta_{esc})
\end{align}
where $\lambda$ is the ratio between the gravitational and thermal energies.
\begin{equation}
\lambda(r)=\frac{GMm}{k_{B}T_{exo}r}=\frac{v_{esc}(r)^2}{U^2}
\label{lambda}
\end{equation}
with $r$ the distance from the center of the body, $v_{esc}(r)$ the escaping velocity, $U$ the most probable velocity for the maxwellian distribution, $G$ the gravitational constant, $M$ the mass of the planet or the satellite and $T_{exo}$ the temperature at the exobase considered constant in the exosphere.

The radiation pressure disturbs the ellipses or hyperbolas described by these particles. The resonant scattering of solar photons leads to a total momentum transfer from the photon to the atom or molecule. In the non-relativistic case, assuming an isotropic reemission of the solar photon, this one is absorbed in the Sun direction and scattered with the same probability in all directions. For a sufficient flux of photons in the absorption wavelength range, the reemission in average does not induce any momentum transfer from the atom/molecule to the photon. The differential of momentum between before and after the scattering each second imparts a force, the radiation pressure. \citet{Bishop1989} proposed to take into account this effect on the structure of planetary exospheres. In particular, they highlighted analytically the ``tail" phenomenon: the density for atomic Hydrogen is higher in the nightside direction than in the dayside direction, as observed for the first time by OGO-5 (\citet{Thomas1972}, \citet{Bertaux1973}).

This problem is similar to so-called Stark effect \citep{Stark1914}: the effect of a constant electric field on the atomic Hydrogen's electron. Its study shows it can be transposed to celestial mechanics in order to describe the orbits of artificial and natural satellites in the perturbed two-body problem. A first but incomplete work was performed by \citet{Bishop1989}. They focused on the density profiles along the Sun-planet axis: in the velocity phase space, the problem is thus only 2D (one component of the angular momentum is null, $p_\phi$, and thus the problem takes place on a hyperplane in the 3D-velocity phase space). They determined the density profiles for bounded trajectories (only ballistic particles, neither escaping nor satellite particles) for atomic Hydrogen along the Sun-planet axis, on the dayside and the nightside, for Earth, Venus, Mars or for Sodium at Mercury.

In this work, we generalize the formalism developed by \citet{Bishop1989} to the whole exosphere (3D case) and highlight several phenomena. Our study is based on Beth et al.(2015a), where we detailed the dynamical aspects induced by the radiation pressure on the trajectories of exospheric particles. We now present the implications on exospheric density profiles, local time asymmetries as well as a specific study of the particles with satellite orbits. We will briefly describe the formalism used in section \ref{model}, before we derive the neutrals density in section \ref{density profiles}, and present the results in section \ref{results} and conclude in section \ref{conclusions}.

\section{Model}\label{model}

For this work, we decide to study the effect of the radiation pressure on atomic Hydrogen in particular. We model the radiation pressure by a constant acceleration $a$ coming from the Sun. As previously defined by \citet{Bishop1991}, this acceleration depends on the line center solar Lyman-$\alpha$ flux $f_0$, in $10^{11}$ photons.cm$^{-2}$.s$^{-1}$.{\AA}$^{-1}$:
\begin{equation}
a=0.1774\ f_0\ (\text{cm.s}^{-2})
\end{equation}

This problem is similar to the classical Stark effect \citep{Stark1914}: a constant electric field (here the radiation pressure) applied on an electron (here an Hydrogen atom) attached to a proton (here the planet). Both systems are equivalent because the force applied by the proton (the planet) on the electron (the Hydrogen atom), the electrostatic force, varies in $r^{-2}$ as the gravitational force from the planet on the Hydrogen atom. Thus, we adopt the same formalism as \citet{Sommerfeld1934} adopting the parabolic coordinates. We use the transformation:
\begin{equation}
\begin{array}{rcccl}
u&=&r+x&=&r(1+\cos\theta)\\
w&=&r-x&=&r(1-\cos\theta)
\end{array}
\end{equation}
where $x$ is the sunward coordinate and $\theta$ the angle with respect to the Sun-planet axis. Along the Sun-planet axis, $w$ is null in the sunward direction whereas $u$ is null in the nightside direction. Consequently, the Hamiltonian becomes:
\begin{equation}
\begin{array}{l}
\mathcal{H}(u,w,p_u,p_w,p_\phi)\\\\
=\dfrac{2up^{2}_u+2wp^{2}_{w}}{m(u+w)}+\dfrac{p^{2}_{\phi}}{2muw}-\dfrac{2GMm}{u+w}+ma\dfrac{u-w}{2}
\label{hamiltonuw}
\end{array}
\end{equation}
independent of $t$ and $\phi$. $p_u$, $p_w$ and $p_\phi$ are the conjugate momenta, $GM$ the standard gravitational parameter of the planet and $m$ the mass of the species.

According to canonical relations, we have:
\begin{equation}
\begin{array}{lcl}
\displaystyle{p_{u}}&=&\dfrac{m(u+w)}{4u}\dfrac{\mathrm{d} u}{\mathrm{d}t} \\\\
\displaystyle{p_{w}}&=&\dfrac{m(u+w)}{4w}\dfrac{\mathrm{d} w}{\mathrm{d}t} \\\\
\displaystyle{p_{\phi}}&=&m uw \dfrac{\mathrm{d}\phi}{\mathrm{d}t} 
\end{array}
\label{pupw}
\end{equation}

We do not assume $p_\phi=0$  as \citet{Bishop1989} did: their study is restricted to the Sun-planet axis where either $u=0$ or $w=0$.

As shown by \citet{Bishop1989}, the problem has three constants of the motion: $\mathcal{H}$, $p_\phi$ and $A$ defined as
\begin{equation}
\begin{array}{rcl}
A&=&2muE-4up^{2}_u-\dfrac{p^{2}_{\phi}}{u}-m^2au^2+2GMm^2\\
&=&-2mwE+4wp^{2}_w+\dfrac{p^{2}_{\phi}}{w}-m^2aw^2-2GMm^2 
\end{array}
\label{A}
\end{equation}

As function of these three constants, we can rewrite the conjugate momenta:
\begin{equation}
\begin{array}{rcl}
p_u&=&\pm\sqrt{\dfrac{-P_3(u)}{4u^2}}\\\\
p_w&=&\pm\sqrt{\dfrac{Q_3(w)}{4w^2}}\\\\
\end{array}
\label{pupw}
\end{equation}
with
\begin{equation}
\begin{array}{rcl}
P_3(u)&=&mau^3-2mEu^2-(2GMm^2-A)u+p^{2}_{\phi}\\\\
Q_3(w)&=&maw^3+2mEw^2+(2GMm^2+A)w-p^{2}_{\phi}\\\\
\end{array}
\end{equation}

We can remark here that for $p_\phi=0$, $0$ is a simple root of $P_3$ and $Q_3$.

Based on the formalism detailed in Beth et al. (2015a), we use dimensionless quantities and the same annotations:
\begin{equation}
\begin{array}{ccl}
\mathcal{E}&=&\dfrac{2UP^{2}_U+2WP^{2}_W}{U+W}+\dfrac{P^{2}_{\phi}}{2UW}-\dfrac{2\lambda_a}{U+W}+\dfrac{\lambda_a}{2}(U-W)
\end{array}
\label{equation357}
\end{equation}
\begin{equation}
\begin{array}{ccl}
\mathcal{A}&=&2\mathcal{E}U-4UP^{2}_{U}-\dfrac{P^{2}_{\phi}}{U}+2\lambda_a-\lambda_a U^2\\\\
&=&-2\mathcal{E}W+4WP^{2}_{W}+\dfrac{P^{2}_{\phi}}{W}-2\lambda_a-\lambda_a W^2
\end{array}
\label{equation358}
\end{equation}
\newline
\begin{equation}
\begin{array}{ccl}
P_{3}(U)&=&\lambda_a U^3-2\mathcal{E}U^2+(\mathcal{A}-2\lambda_a)U+P^{2}_{\phi}\\
&=&\lambda_a(U-U_{0})(U-U_{-})(U-U_{+})\\\\
Q_{3}(W)&=&\lambda_a W^3+2\mathcal{E}W^2+(\mathcal{A}+2\lambda_a)W-P^{2}_{\phi}\\
&=&\lambda_a(W-W_{-})(W-W_{+})(W-W_{0})
\end{array}
\label{equation359}
\end{equation}
\newline
with $U_0$, $U_-$, $U_+$ and $W_0$ real roots such as $U_0<0<U_-<U_+$; $W_-$ and $W_+$ can be real (then $W_-<W_+<W_0$) or complex conjugates. For each polynom, one root can be $0$ if $P_\phi=0$. We define $\lambda_a$ as:
\begin{equation}
\lambda_a=\dfrac{\sqrt{GMa}m}{k_{B}T}=\dfrac{GMm}{k_B T R_{pressure}}=\lambda(R_{pressure})
\end{equation}
the Jeans parameter at the distance $R_{pressure}=\sqrt{GM/a}$.

\section{Calculation of exospheric densities}\label{density profiles}

The density in a gas is given from the distribution function $f$ by:
\begin{equation}
n(\vec{r})=\int\!f(\vec{r},\vec{v})\, \mathrm{d}^3\vec{v}
\label{density}
\end{equation}

The bounds of integration depend on the type of particles. As for the so-called \citet{Chamberlain1963} model, we distinguish three types of particles: ballistic, satellite and escaping particles. As detailed in the next section (and summarized in table \ref{table31}), the trajectory of these particles are not conics at all but they keep a part of their definition in this problem: the ballistic particles cross twice the exobase, the satellite particles do not cross the exobase and do not escape, and the escaping particles cross the exobase once before they escape. As shown by Beth et al. (2015a), each type of particles must respect conditions about the roots of $P_3$ and $Q_3$, summarized in table \ref{table31}.

\begin{table}[!h]
\begin{center}
\begin{tabular}{|l||c|c|rr|}
\hline
{\rule[-1.2ex]{0ex}{4ex}type of trajectory}&ballistic&satellite&\multicolumn{2}{c|}{escaping} \\
\hline
\hline
\rule[-1.2ex]{0ex}{4ex}positive real roots of $P_3$    &     2                                   &     2                              & \multicolumn{2}{c|}{$2$} \\
\hline
\rule[-1.2ex]{0ex}{4ex}positive real roots of $Q_3$    &      3                                   &     3                                & 1 & 3 \\
\hline

\rule[-2ex]{0ex}{6ex}condition on $\dfrac{r_{exo}}{R_{pressure}}$              &        $>\dfrac{U_{-}+W_{-}}{2}$&    $<\dfrac{U_{-}+W_{-}}{2}$       &\multicolumn{2}{c|}{$>\dfrac{U_{-}+W_{0}}{2}$}\\
\hline
\rule[-1.2ex]{0ex}{4ex}Necessary and sufficient & \multirow{2}{*}{Yes}& \multirow{2}{*}{Yes}& \multicolumn{2}{c|}{\multirow{2}{*}{No, need for tracking}}\\
conditions?&&&&\\
\hline
\end{tabular}
\end{center}
\caption{Conditions on the roots of the polynoms and on the exobase location to define the ballistic, satellite and escaping trajectories}
\label{table31}
\end{table}

\subsection{Reminder about the previous analytical work by \citet{Bishop1989}}

Before performing the density profile calculation, we specify here the main differences with the initial work by \citet{Bishop1989}.

\citet{Bishop1989} performed 1D calculations along the Sun-planet axis only, implying some assumptions on the trajectories of the particles (described below). We explain here rigorously why their study and ours are different and complementary.

So that a particle crosses the Sun-planet axis, $U$ or $W$ will be null once and thus $P_\phi$ too (cf. eq \ref{pupw}) because it is independent of the time. Thus, $P_3$ and $Q_3$ have $0$ as common root. For bounded trajectories, $W_-$, $W_+$ and $W_0$ are real positive and $0<W_-<W_+<W_0$ thus $W_-=0$. The question now is: which root of $P_3$ is null? As explained further, $U_0+U_++U_-=-W_+-W_0<0$ and $U_+$ is strictly positive. Then, between $U_-$ and $U_0$, one is negative, the other one is null thus $U_0<0$ and $U_-=0$.

The $(U,W)$-motion of a particle crossing the Sun planet axis is $(U,W)\in[0;U_+]\times[0;W_+]$. According to the Poincaré recurrence theorem as used by Beth et al. (2015a), these particles will necessarily pass as close as we want to the $(0,0)$ position. Consequently, as explained by \cite{Bishop1989} for the planar motion ($P_{\phi}=0$), all bounded trajectories cross the exobase. These authors considered the particles orbiting severa times around the planet as satellite particles with a finite lifetime. We will show in the following sections for our 3D-case, that pure satellite particles exist which never cross the exobase (see section \ref{satpart}). Finally, our integrals here will be calculated in another way compared with \citet{Bishop1989} since $P_\phi$ can vary. Although our model cannot take into account the singularity $P_\phi=0$, this is not a problem for the integration (for the numerical evaluation, we just avoid this value).

\subsection{Densities of ballistic particles}

We will first focus on the densities of ballistic particles. These particles are trapped by the potential that includes both the gravity and the radiation pressure. Consequently, the parabolic coordinates $U$ and $W$ must have finite values and belong to a closed interval from $\mathbb{R}^+$. On the first hand, there is no further constraint for $U$-values, since the $U$-motion is always limited by two paraboloids defined by $U_-$ and $U_+$ (see section 2.4 in Beth et al. (2015a)). On the other side, the $W$-motion must respect the following condition to allow for ballistic particles: the roots $W_-$ and $W_+$ must be real positive (otherwise, they can be complex or real negative). This implies a condition on the energy $\mathcal{E}$ according to equation \ref{equation359}:
\begin{equation}
\lambda_a(W_0+W_-+W_+)=-2\mathcal{E}>0
\label{energy}
\end{equation}

\begin{figure}
\includegraphics[height=.35\linewidth]{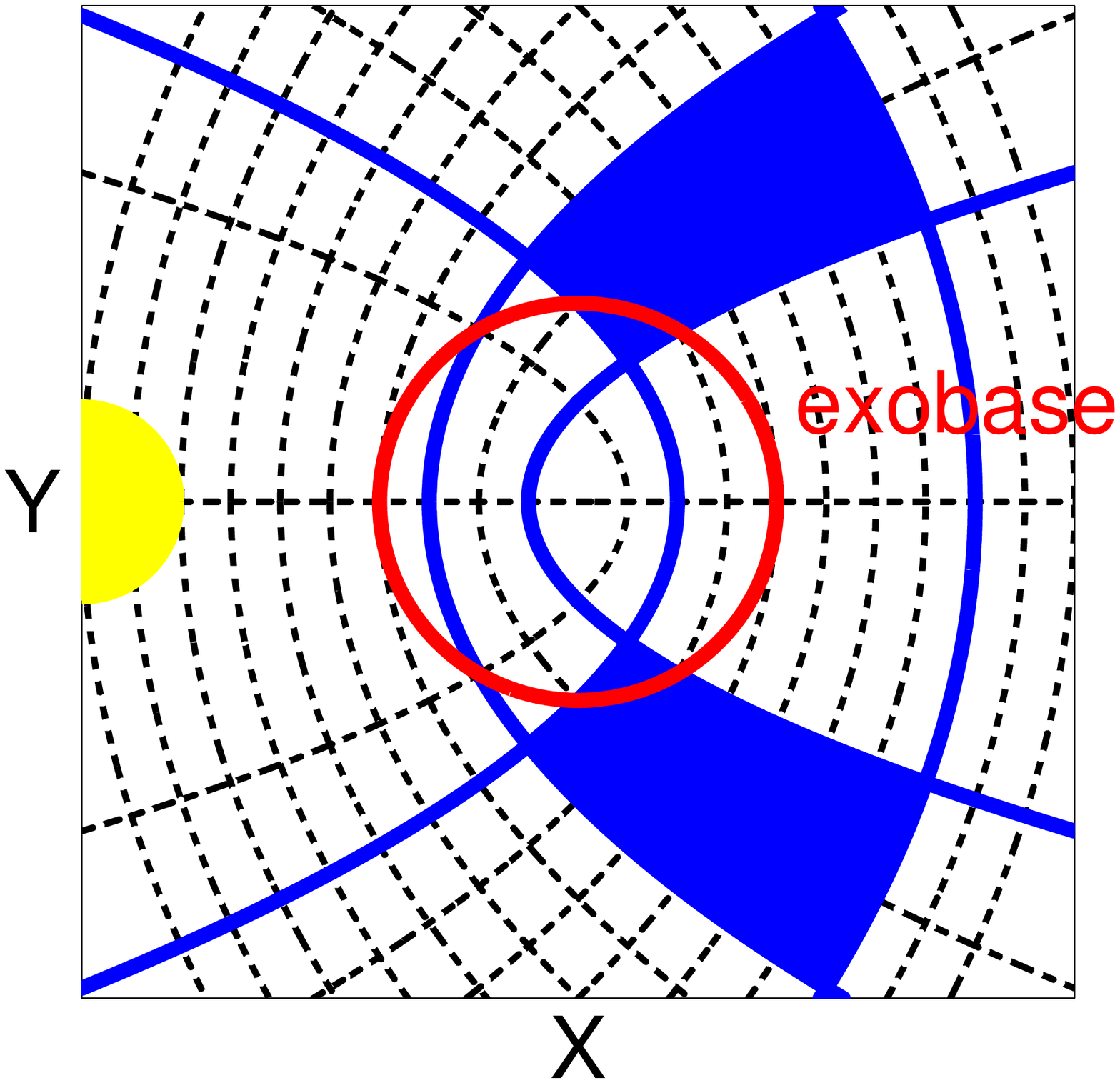}\includegraphics[height=.35\linewidth]{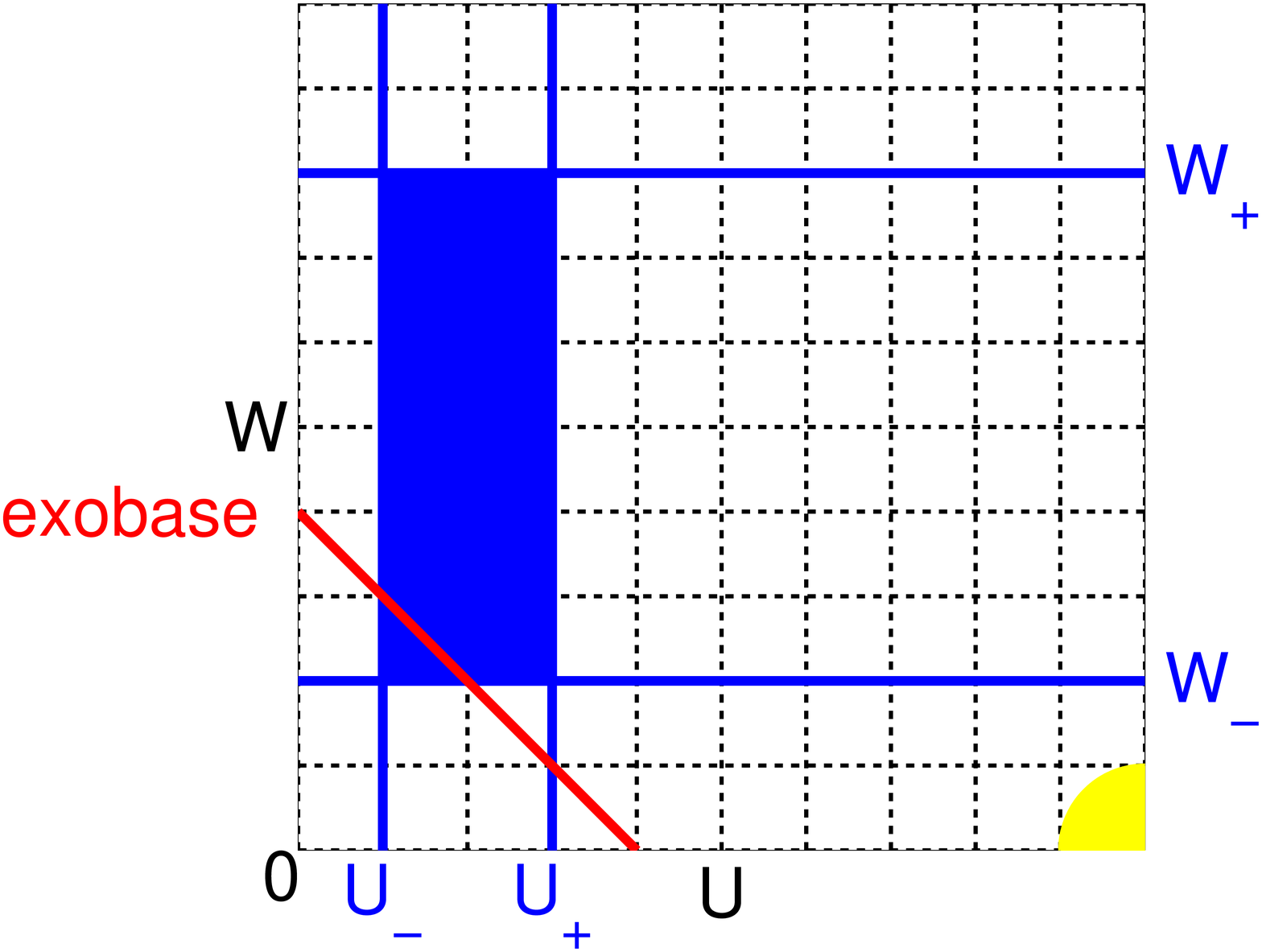}
\includegraphics[height=.35\linewidth]{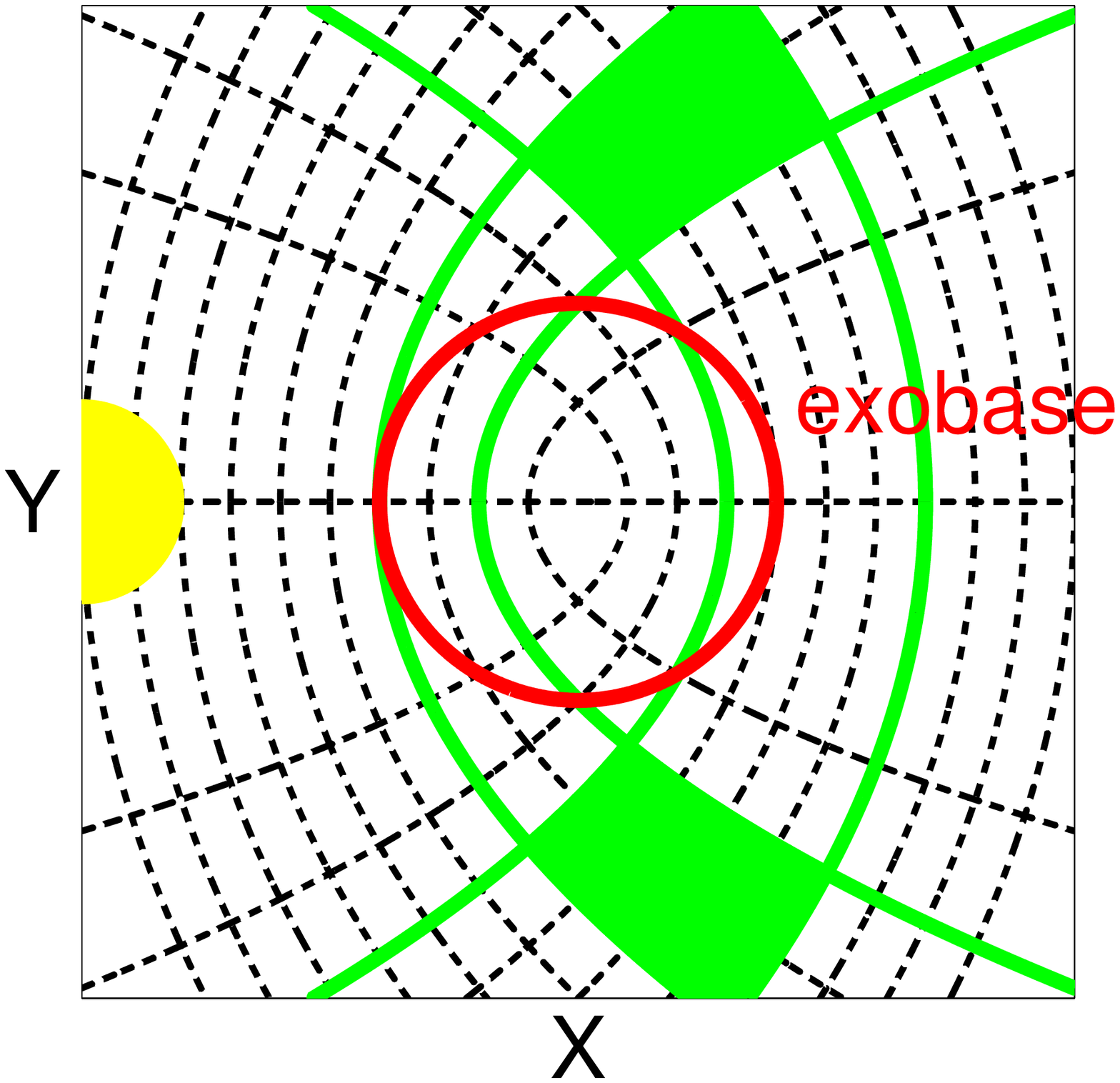}\includegraphics[height=.35\linewidth]{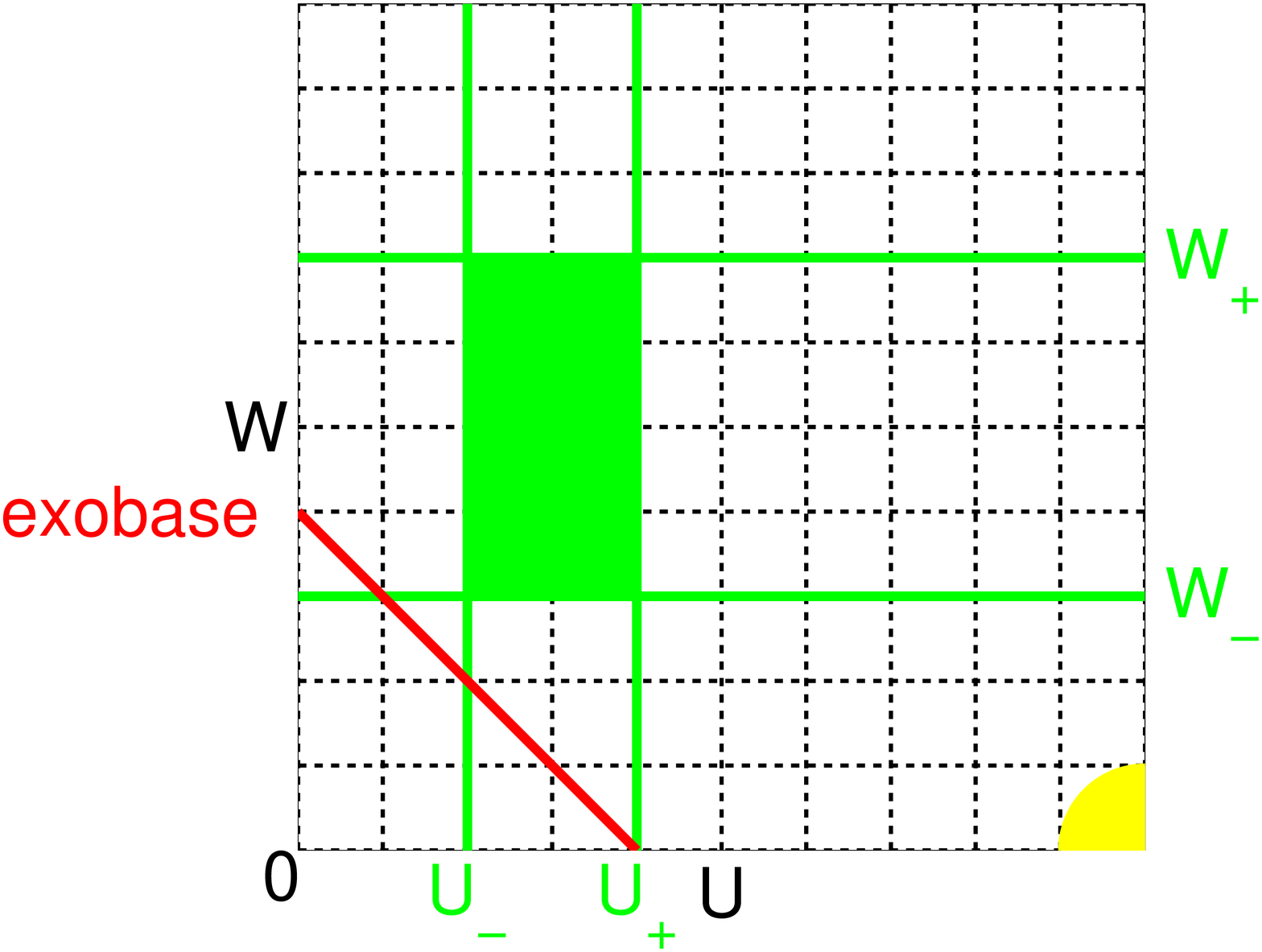}
\caption{Representations of areas where ballistic (blue) and satellite (green) particles evolve in the usual coordinate system (left panel) and in the $(U,W)$ plane. To pass from the left grid to the right one, we apply the transformation $(x,y)\rightarrow(u=\sqrt{x^2+y^2}+x,w=\sqrt{x^2+y^2}-x)$.}
\label{restricted}
\end{figure}

As explained by Beth et al. (2015a), the $U$-motion does not have the same period as the $W$-motion or the $\phi$-motion. In the $(U,W)$-plane, the trajectory fills the whole square $[U_-;U_+]\times[W_-;W_+]$ (see Beth et al. (2015a), fig. \ref{restricted}). If a particle belongs to this square, after a certain time, the particle will pass as close as we want to any position chosen in this square (except for specific periodic cases, cf. \cite{Biscani2014}, occurring only if we are looking for them). Thus, if a part of the exobase surface belongs to this square, the particle in this square will necessarily cross several times the exobase (at the distance $r_{exo}$). We will thus be able to call it a ballistic particle. By definition $u+w=(U+W)R_{pressure}=2r$, and the nearest distance from the center of the planet where the particle can pass is $r_-=(U_-+W_-)R_{pressure}/2$. If $W_-$ is real positive, then $r_-$ allow us to define the type of this particle: if $r_-<r_{exo}$ the particle is ballistic, if $r_->r_{exo}$ the particle is satellite.

In summary, the bounds of integration needed to calculate the density of ballistic particles from equation \ref{density} is the part of the velocity space where $W_-$ and $W_+$ are real positive and $r_-<r_{exo}$.

A last parameter must be determined: what is the distribution function to be used? The calculation of the ballistic density (detailed below) is based on the Liouville theorem, which uses the conservation of the distribution function along a dynamical trajectory. Since the ballistic particles cross the exobase, i.e. the external limit of the collisional atmosphere, one can reasonably assume (as in \citet{Chamberlain1963}) that the disitribution function is maxwellian at the exobase.
\newline

\subsection{Description of the algorithm for ballistic particles}

\textbf{First step}
We choose a specific location in the exosphere and we thus fix the values of $U$ and $W$.\newline

\textbf{Second step}
Then, we scan all possibilities for the velocity vector. For each 3-tuple (three components of the velocity vector), we calculate the corresponding three constants of the motion ($\mathcal{E}$, $\mathcal{A}$, $P_\phi$).\newline

\textbf{Third step}
For each tuple  ($\mathcal{E}$, $\mathcal{A}$, $P_\phi$), we calculate the three roots of $P_3$ and $Q_3$.
\newline

\textbf{Fourth step}
We test if the three roots of $Q_3$ are real and positive.
\newline

\textbf{Fifth step}
We calculate $U_-/2+W_-/2$. If this value is below $r_{exo}/R_{pressure}$ then this is a bounded trajectory crossing the exobase, that corresponds to ballistic particles. Thus, the Liouville theorem can be applied.\newline

\textbf{Sixth step}
We integrate the velocity distribution function and derive the density of ballistic particles.\newline

In order to calculate the ballistic density $n_{bal}$, we propose to define a partition function $\zeta_{bal}(r,\theta)$ that is the generalization, when the radiation pressure effect is included, of the 1D partition function defined by \cite{Chamberlain1963}:
\begin{equation}
n_{bal}(r,\theta)=\underbrace{n_{exo}\exp(\lambda(r)-\lambda_c)}_{\text{barometric law}}\exp\left(-\lambda_a\dfrac{r-r_{exo}}{R_{pressure}}\cos\theta\right)\zeta_{bal}(r,\theta)
\end{equation}
with $n_{exo}$ the exobase density, $\theta$ the angle with the respect to the Sun-planet axis and $r$ the distance from the planet. By definition, for a Maxwellian distribution,
\begin{equation}
\zeta_{\text{bal}}(r,\theta)=\dfrac{\displaystyle \int_{\text{bal}} \!  \exp\left(-\dfrac{p^2}{2mk_B T}\right) \, \mathrm{d}^3\vec{p}}{\displaystyle \int \! \exp\left(-\dfrac{p^2}{2mk_B T}\right) \, \mathrm{d}^3\vec{p}}
\end{equation}

We choose to work with dimensionless quantities and the $(U,W)$ coordinates. Operating some transformations and with the correct Jacobian, we obtain:
\begin{equation}
\begin{array}{ll}
\zeta_{\text{bal}}(U,W)=\\
\dfrac{1}{(2\pi)^{3/2}}\dfrac{4}{U+W}\displaystyle \int_{bal} \!  \exp\left(-\dfrac{2UP_{U}^2+2WP_{W}^2}{U+W}-\dfrac{P_{\phi}^2}{2UW}\right) \, \mathrm{d}P_U\,\mathrm{d}P_W\,\mathrm{d}P_\phi
\end{array}
\label{integral}
\end{equation}

The formula \ref{integral} is only available for $U\neq0$ or $W\neq0$. Indeed, if $U=0$ or $W=0$, then $P_{\phi}=0$ and this integral cannot be performed. Thus, our formulation is only available for the whole exosphere except along the Sun-planet axis, already studied by \cite{Bishop1989}.

As previously mentioned, $\mathcal{E}$ is negative. It represents the contribution of the potential energy and the kinetic energy. Thus, we have the inequality from the equation \ref{equation357}:
\begin{equation}
\dfrac{2\lambda_a}{U+W}-\dfrac{\lambda_a}{2}(U-W)>\dfrac{2UP_{U}^2+2WP_{W}^2}{U+W}+\dfrac{P_{\phi}^2}{2UW}>0
\label{limite}
\end{equation}
The modulus inside the exponential takes finite values. In this case, we choose the following change of coordinates:
\begin{equation}
\begin{array}{ccccl}
X&=&\sqrt{\dfrac{2U}{U+W}}P_{U}&=&R\sin \Theta \cos \Phi\\\\
Y&=&\sqrt{\dfrac{2W}{U+W}}P_{W}&=&R\sin \Theta \sin \Phi\\\\
Z&=&\dfrac{1}{\sqrt{2UW}}P_{\phi}&=&R \cos \Theta
\label{equation388}
\end{array}
\end{equation}

\begin{equation}
\begin{array}{rcl}
\zeta_{\text{type}}(U,W)&=&\dfrac{1}{\pi^{3/2}}\displaystyle \int_{\text{bal}} \! \exp(-X^2-Y^2-Z^2) \, \mathrm{d}X\,\mathrm{d}Y\,\mathrm{d}Z\\\\
&=&\dfrac{1}{\pi^{3/2}}\displaystyle \int_{\text{bal}}\! R^2 \exp(-R^2) \, \mathrm{d}R\,\mathrm{d}\Theta\,\mathrm{d}\Phi\\\\
\end{array}
\label{densite_integ}
\end{equation}

\begin{figure}[!h]
\centering
\noindent\resizebox{0.8\linewidth}{!}{
\begin{tikzpicture}[auto]
\matrix(m)[matrix of nodes, column sep=3em, row sep=1.5em,align=center,ampersand replacement=\&]{%
{}\&\node [anchor=center,block] (init) {Initialization: parameters at the exobase, $i,j,k=1$ and $N$ points for the method at the position $(U_0,W_0)$};\&{}  \\
{}\& \node [anchor=center,block] (choix1) {Iteration on $i$, $j$, $k$ going from $1$ to $N$. Choice of the nodes $R_i$, $\Theta_j$ and $\Phi_k$ imposed by the Gauss-Legendre method};\&{}\\
{}\&  \node [anchor=center,block] (calcul0) {Calculation of $X$, $Y$ and $Z$ (Eq. \eqref{equation388})};\&{}\\
{}\&  \node [anchor=center,block] (calcul1) {\mbox{Calculation of $P_U$, $P_W$ and $P_{\phi}$} (Eq. \eqref{equation388})};\&{}\\
{}\& \node [anchor=center,block] (calcul2) {Calculation of $\mathcal{E}$, $\mathcal{A}$ \mbox{(Eq. \eqref{equation357} et \eqref{equation358})}};\&{}\\
 \node [anchor=center,block2] (calcul41) {$\mathds{1}_{ballistic}=0$ }; \& \node [anchor=center,decision] (test1) {$W_+$ and $W_-$ real (Cardano's method)};\&{}\\
{}\&\node [anchor=center,block] (calcul3) {Calculation of $U_{-}$, $W_{-}$ and $W_{+}$ (Cardano's or Newton Raphson method)};\&{}\\
{}\&\node [anchor=center,decision] (test2) {\mbox{$W_0<W_{+}$} };\&{}\\
{}\&\node [anchor=center,decision] (test3) {$U_{-}+W_{-}<2r_{exo}/R_{pressure}$};\& \node [anchor=center,block2] (calcul42) {$\mathds{1}_{bal}=1$};\\
};

\draw[->,line width=.2em] (init) -- (choix1);
\draw[->,line width=.2em] (choix1) -- (calcul0);
\draw[->,line width=.2em] (calcul0) -- (calcul1);
\draw[->,line width=.2em] (calcul1) -- (calcul2);
\draw[->,line width=.2em] (calcul2) -- (test1);
\draw[->,line width=.2em] (test1) -- (calcul3);
\draw[->,line width=.2em] (calcul3) -- (test2);
\draw[->,line width=.2em] (test1) -- node{Yes}(calcul3);
\draw[->,line width=.2em] (test3) -- node{Yes}(calcul42);

\draw[->,line width=.2em] (test1) -- node[above]{No}(calcul41);
\draw[->,line width=.2em] (test2) -| node[pos=0.25,above]{No}(calcul41);
\draw[->,line width=.2em] (test3) -| node[pos=0.25,above]{No}(calcul41);
\draw[->,line width=.2em] (calcul41) |- node[pos=0.75,above]{Increment}(choix1);
\draw[->,line width=.2em] (calcul42) |- node[pos=0.75,above]{Increment}(choix1);
\draw[->,line width=.2em] (test2) -- node{Yes}(test3);

\end{tikzpicture}

}
\caption{Algorithm to determine the value, $0$ or $1$, of the function $\mathds{1}_{ballistic}$ according to the initial conditions. Once this value is known, we use equation \ref{GL}.}
\label{algo1}
\end{figure}
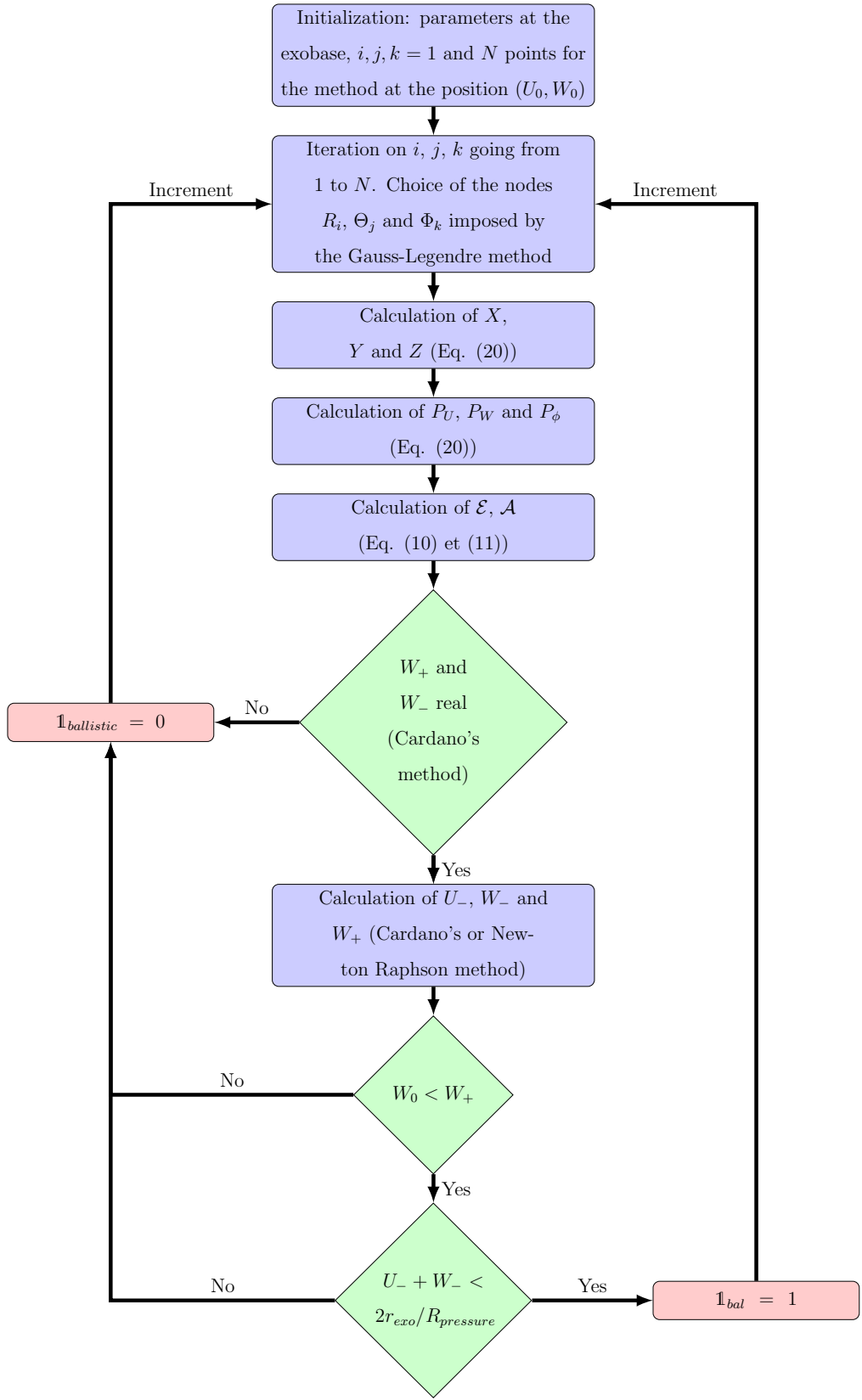

To estimate this integral \ref{densite_integ}, we use the Gauss-Legendre quadrature as:

\begin{equation}
\begin{array}{rcl}
\zeta_{\text{bal}}(U,W)&=&\dfrac{1}{\pi^{3/2}}\displaystyle \int \! \mathds{1}_{\text{bal}}(R,\Theta,\Phi) R^2  \exp(-R^2) \, \mathrm{d}R\,\mathrm{d}\Theta\,\mathrm{d}\Phi\\\\
&\approx&\displaystyle{\dfrac{1}{\pi^{3/2}}\sum_{i=1}^{N}\sum_{j=1}^{N}\sum_{k=1}^{N}  \mathds{1}_{\text{bal}}(R_i,\Theta_j,\Phi_k)  R_{i}^{2}\exp(-R_{i}^{2}) w_i w_j w_k}
\end{array}
\label{GL}
\end{equation}
$N$ is the number of points used for the integration and $\mathds{1}_{\text{bal}}$ is a function taking the value $1$ if the particle is bounded and crosses the exobase with these initial conditions or 0 otherwise, as a rejection sampling (cf. figure \ref{algo1}).

The Gauss-Legendre method is particularly efficient in this case because all bounds of integration are finite: $R$ is finite (see eq. \ref{limite}), $\Theta$ belongs to $[0;\pi]$ and $\Phi$ to $[0;2\pi]$.

\subsection{The escaping and the satellite particles}

Our approach can unfortunately not be applied so easily to calculate the densities of particles with escaping or satellite trajectories.

For escaping particles, the coordinates in the full position-velocity phase space do not guarantee that the particles come from the exobase or not. Indeed, for escaping particles, the volume is opened (no restriction on $\mathcal{E}$). The Poincaré recurrence theorem thus cannot be applied and a particle can have the conditions to be escaping without crossing the exobase (the particle is just passing by the planet): it is necessary to follow the particle along the time as long as it is inside the exopause. If a particle reaches the exopause without crossing the exobase, this is not a escaping particle. We tried to track the particles to know if they cross or not the exobase but we have some time and precision issues: compared with the ballistic particles, we shall compute the trajectory of each particle (problem of time) and integrate the energies until the infinite (problem of precision). Several attempts were performed without convincing results. However, as will be discussed in a future paper, it is possible to calculate the analytical escape flux at the exobase.

The satellite particles cannot exist in our model because this is a collisionless model. We have previously proposed a formalism to estimate their density (\citet{Beth2014}); the trajectories are not closed here (no periodic motions for all bounded cases) and the formalism of this paper cannot be adjusted in this way. However, we will show below (section \ref{satpart}) that satellite particles can exist in the presence of a radiation pressure force, and where these particles are.

\section{Results}\label{results}

\cite{Bishop1989} provided only the ballistic density along the Sun-planet axis, along the dayside and nightside directions. Here, we generalize this approach with a 2D model  (3D if an axisymmetric symmetry is considered) and provide the ballistic density (main component in the lower part of the exosphere) in every direction from the planet. We present here the results for different planetary exospheres such as for Earth, Titan and Mars, but the main features derived (comparison with pure gravity case, asymmetries, exopause) are general results that may be applied to any planet hosting a dense atmosphere influenced by a radiation pressure force.

\subsection{Asymmetries and comparison with Chamberlain profiles}

\begin{figure}[!t]
\centering
\includegraphics[width=0.3\linewidth]{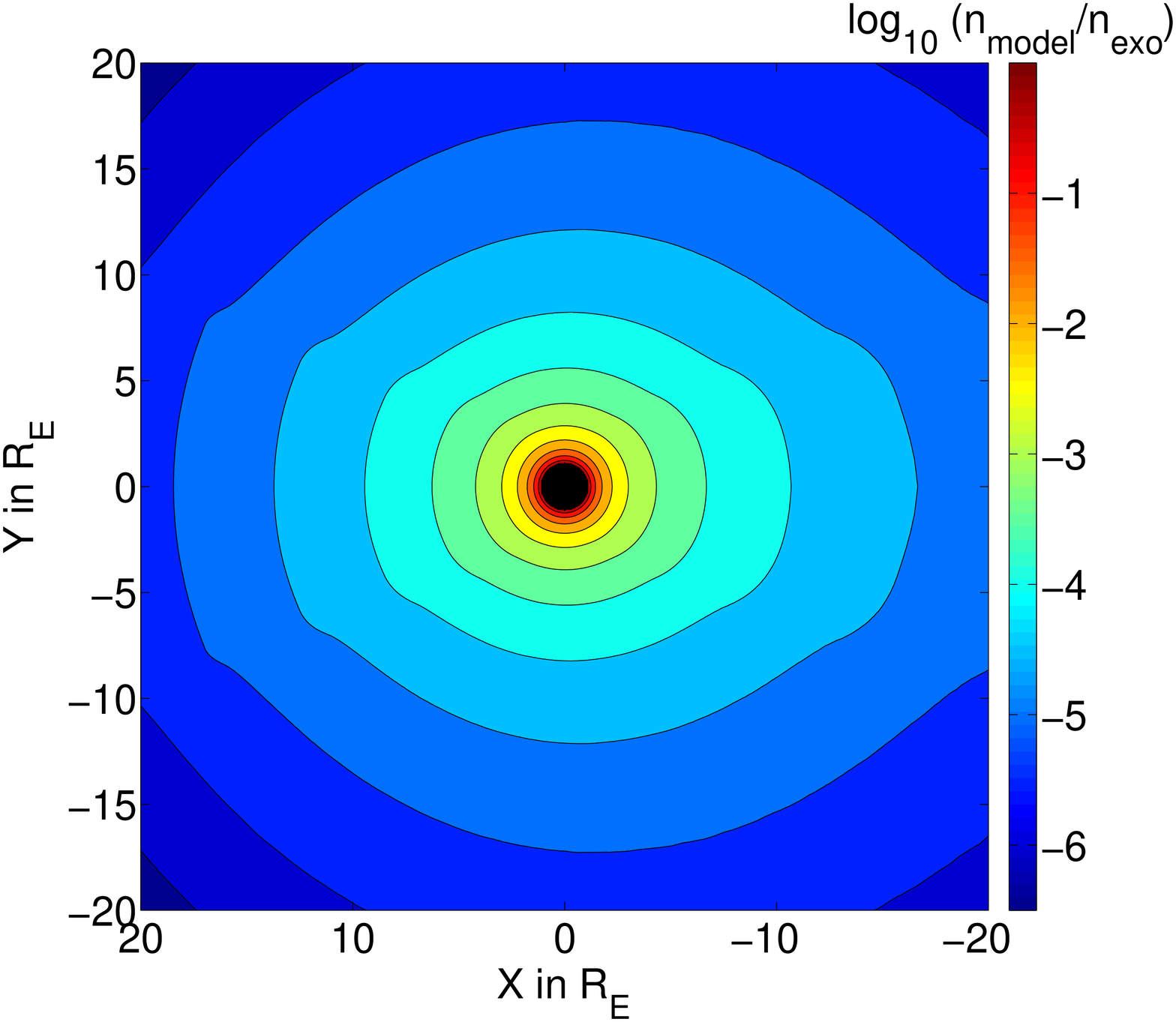}
\includegraphics[width=0.3\linewidth]{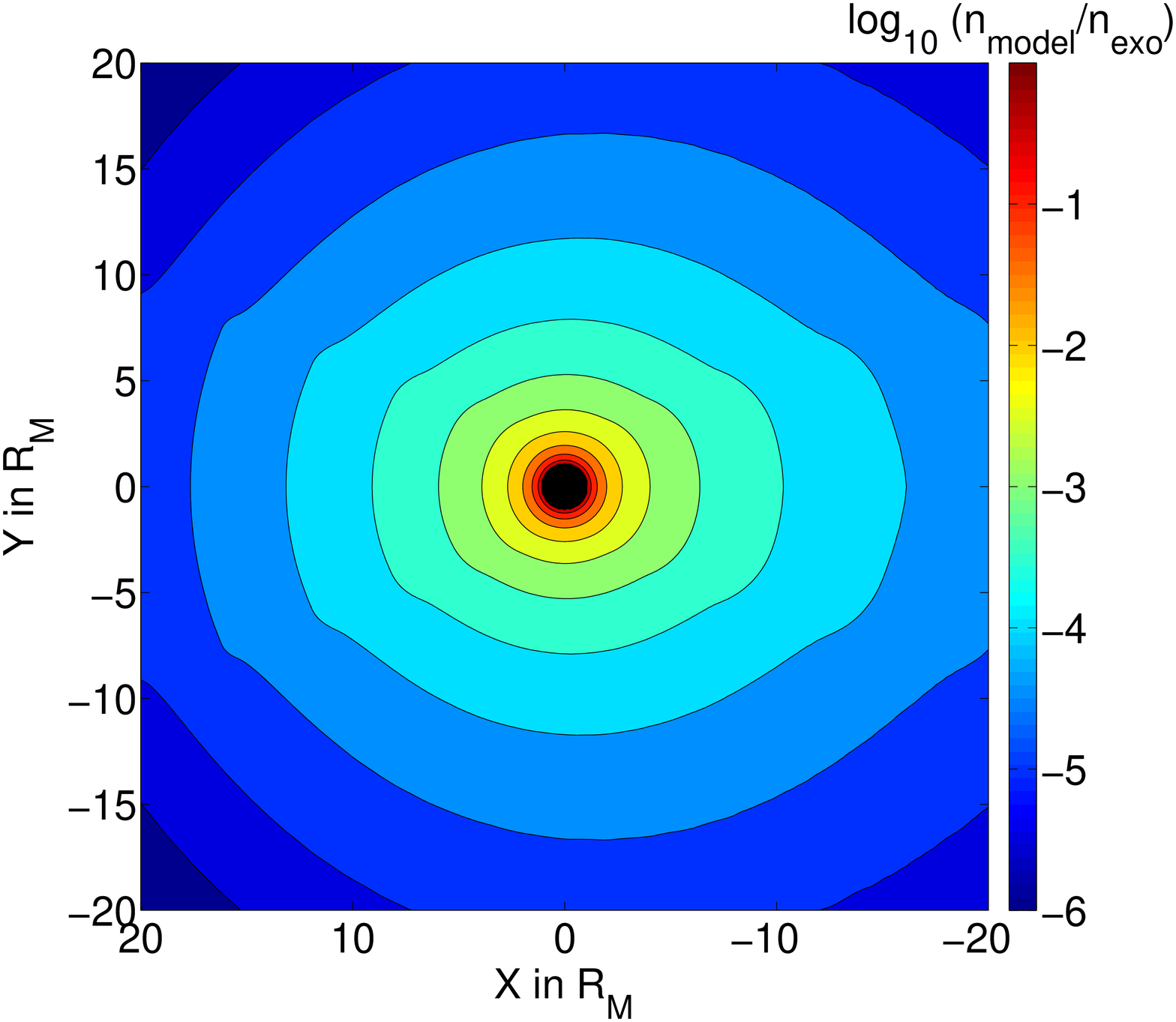}
\includegraphics[width=0.3\linewidth]{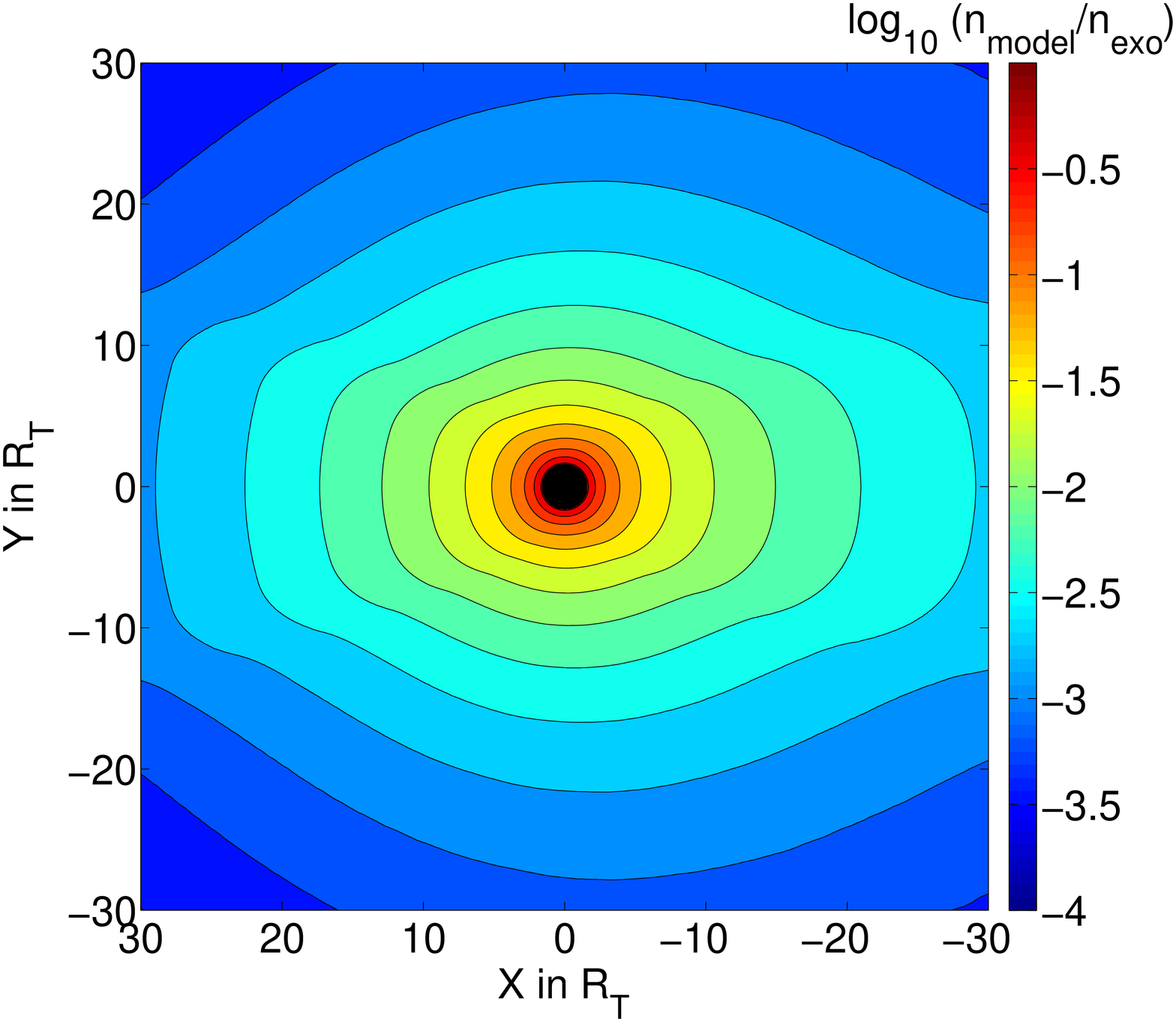}
\includegraphics[width=0.3\linewidth]{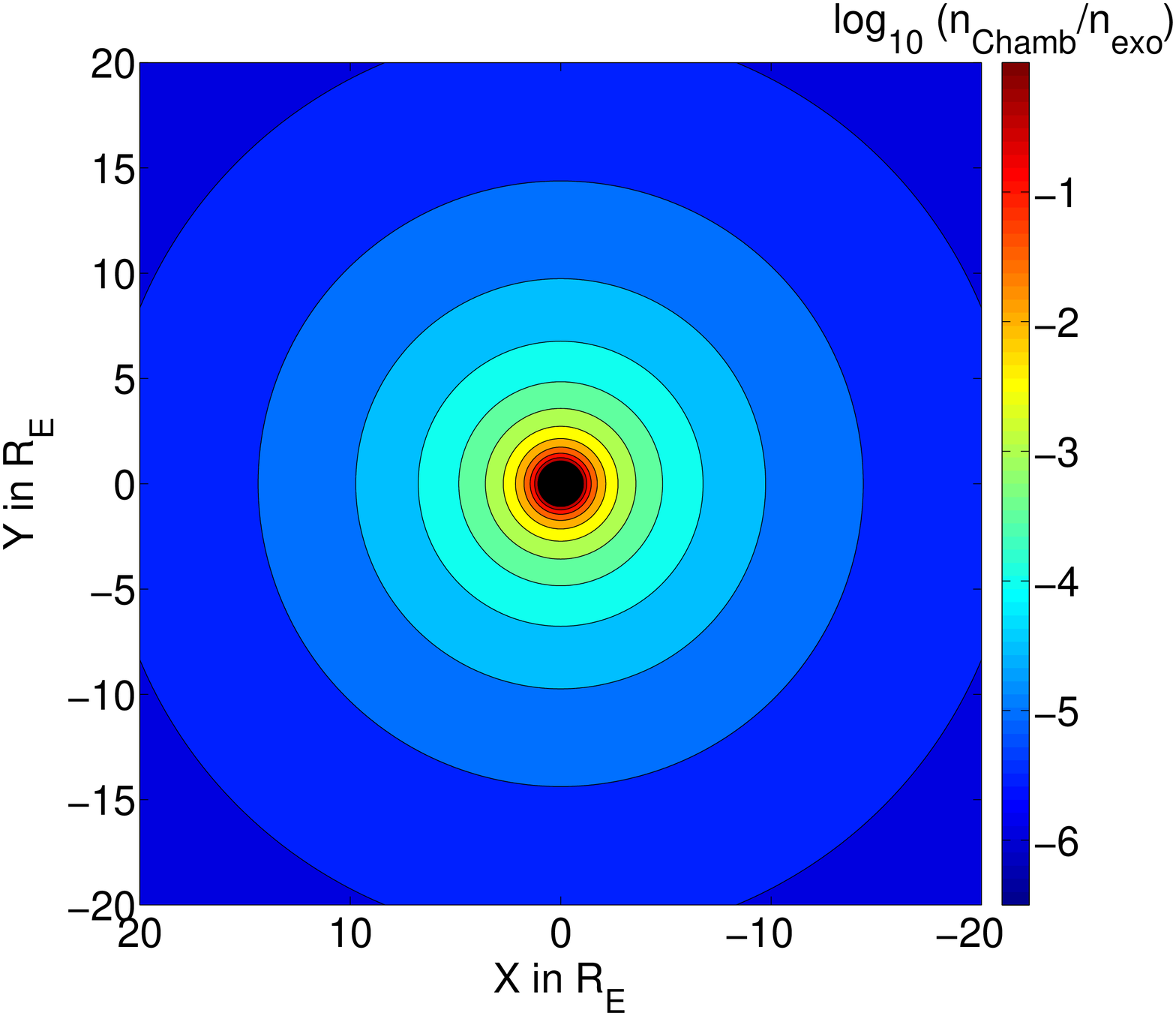}
\includegraphics[width=0.3\linewidth]{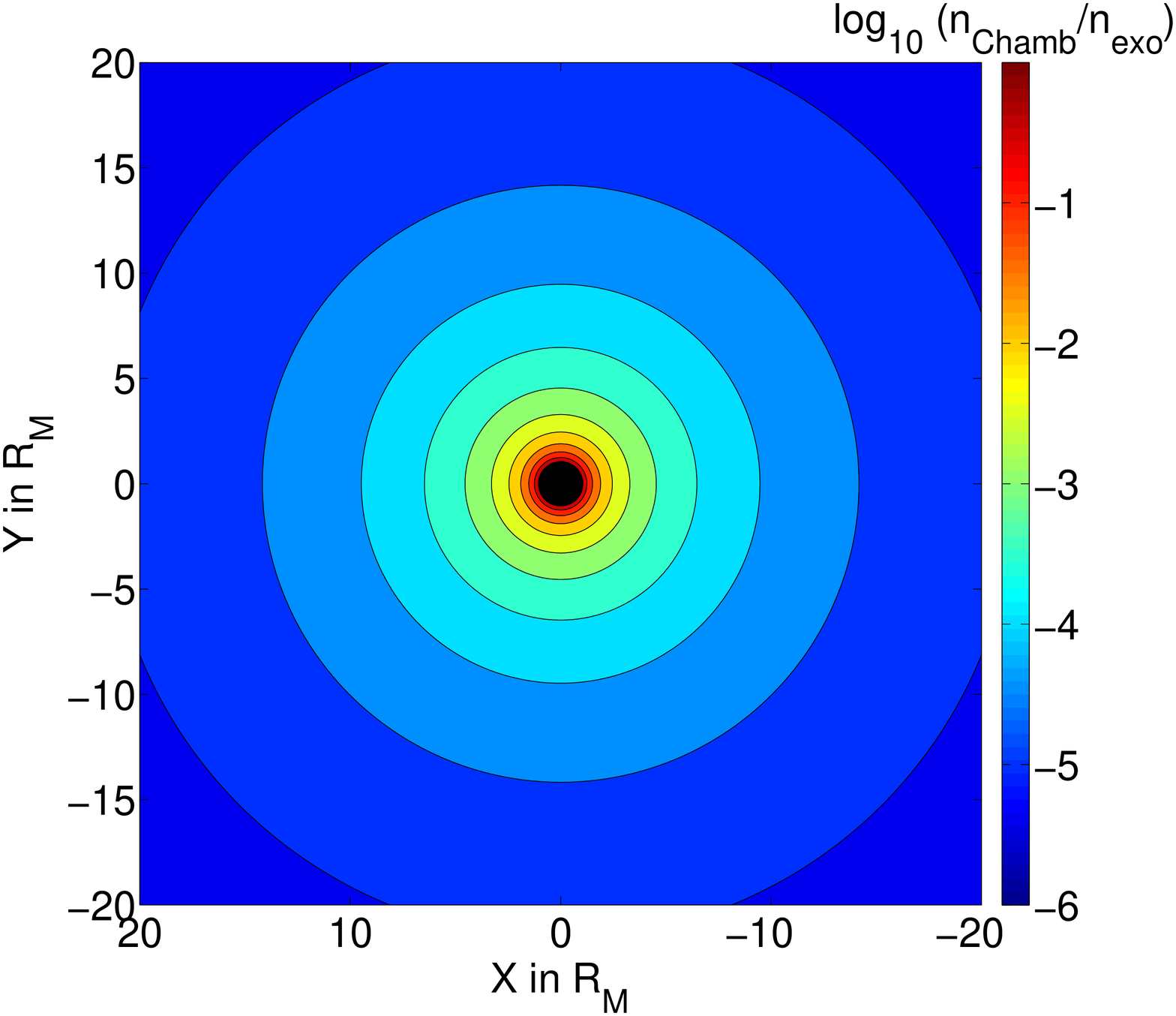}
\includegraphics[width=0.3\linewidth]{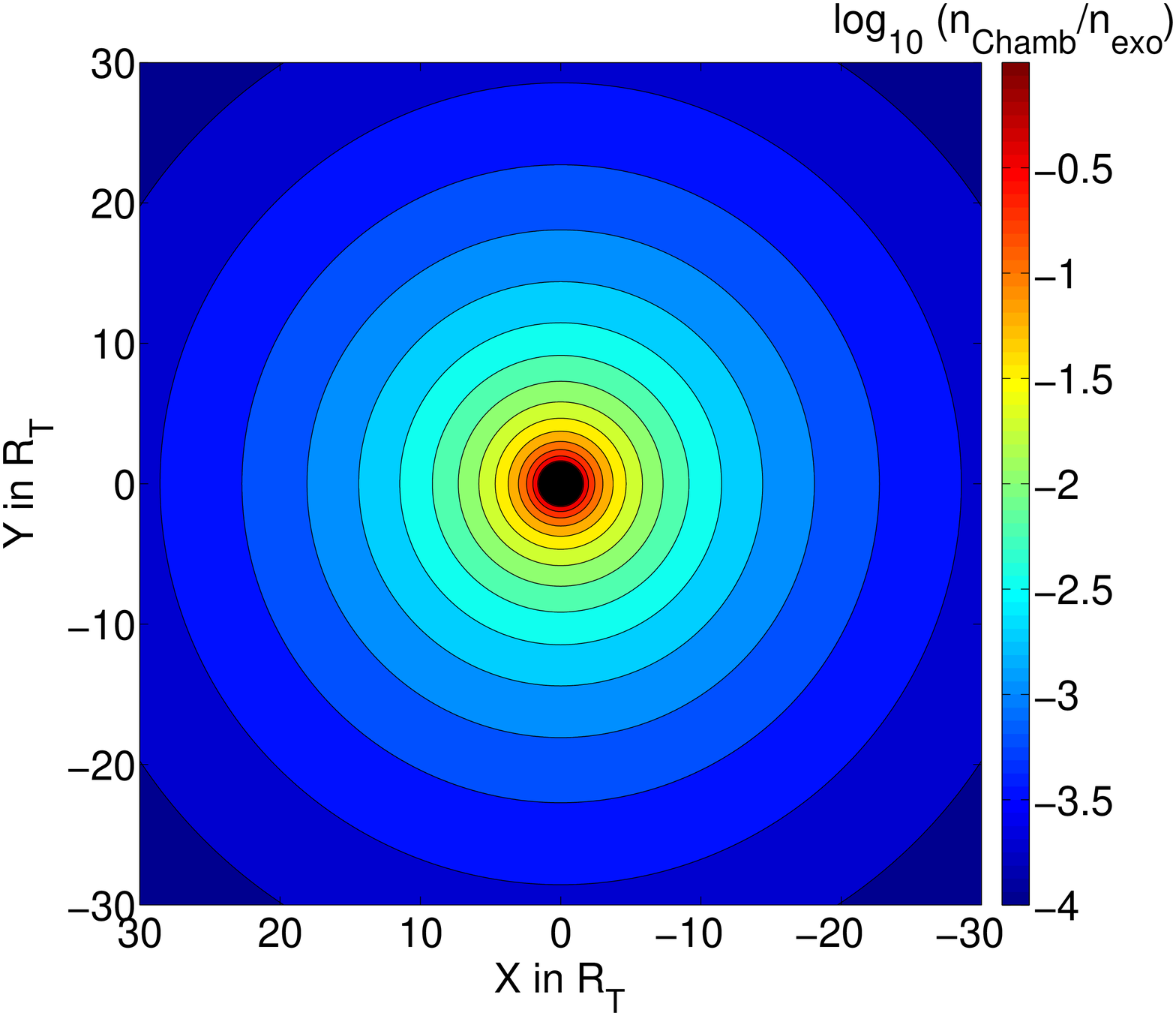}
\includegraphics[width=.3\linewidth]{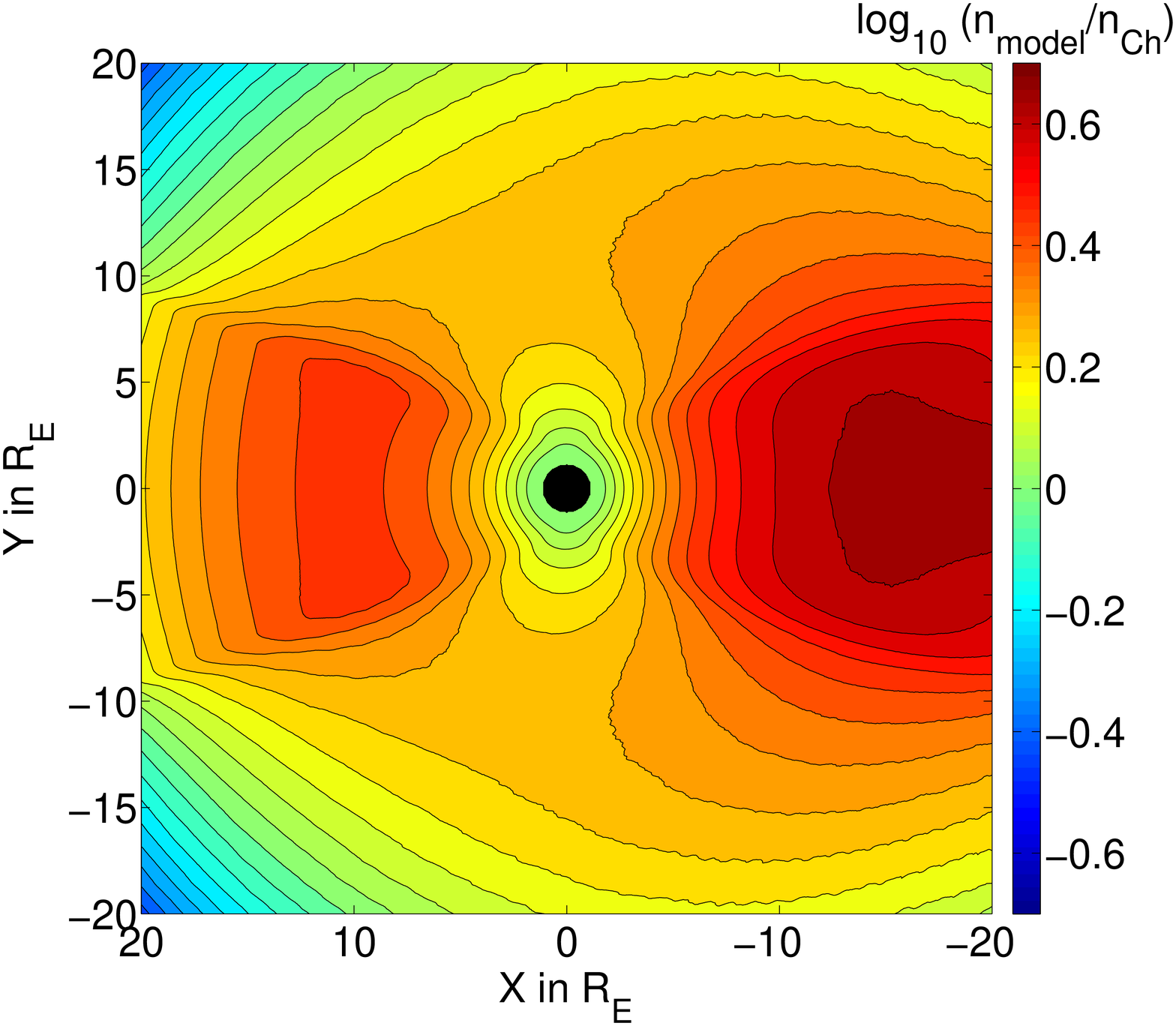}
\includegraphics[width=.3\linewidth]{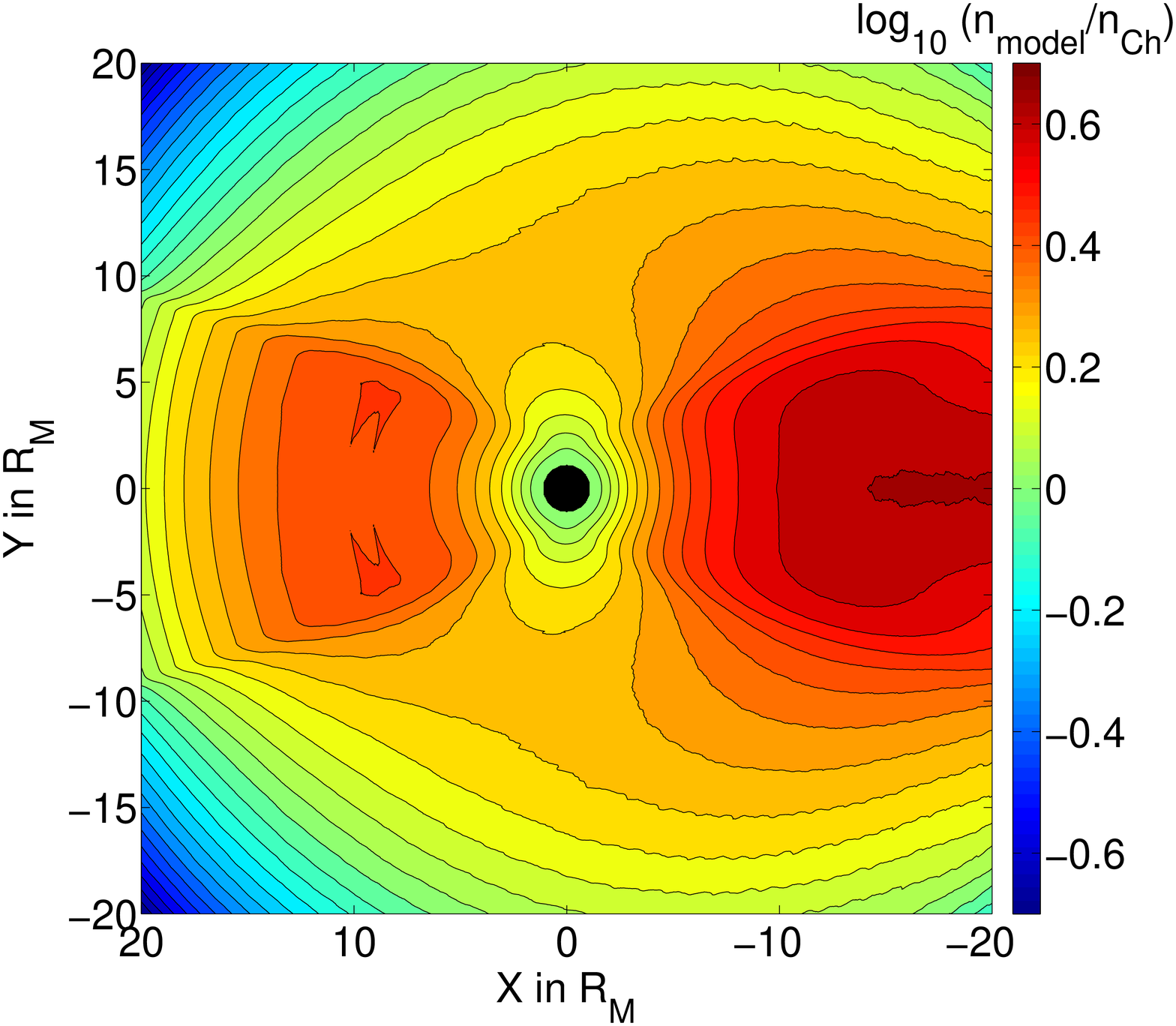}
\includegraphics[width=.3\linewidth]{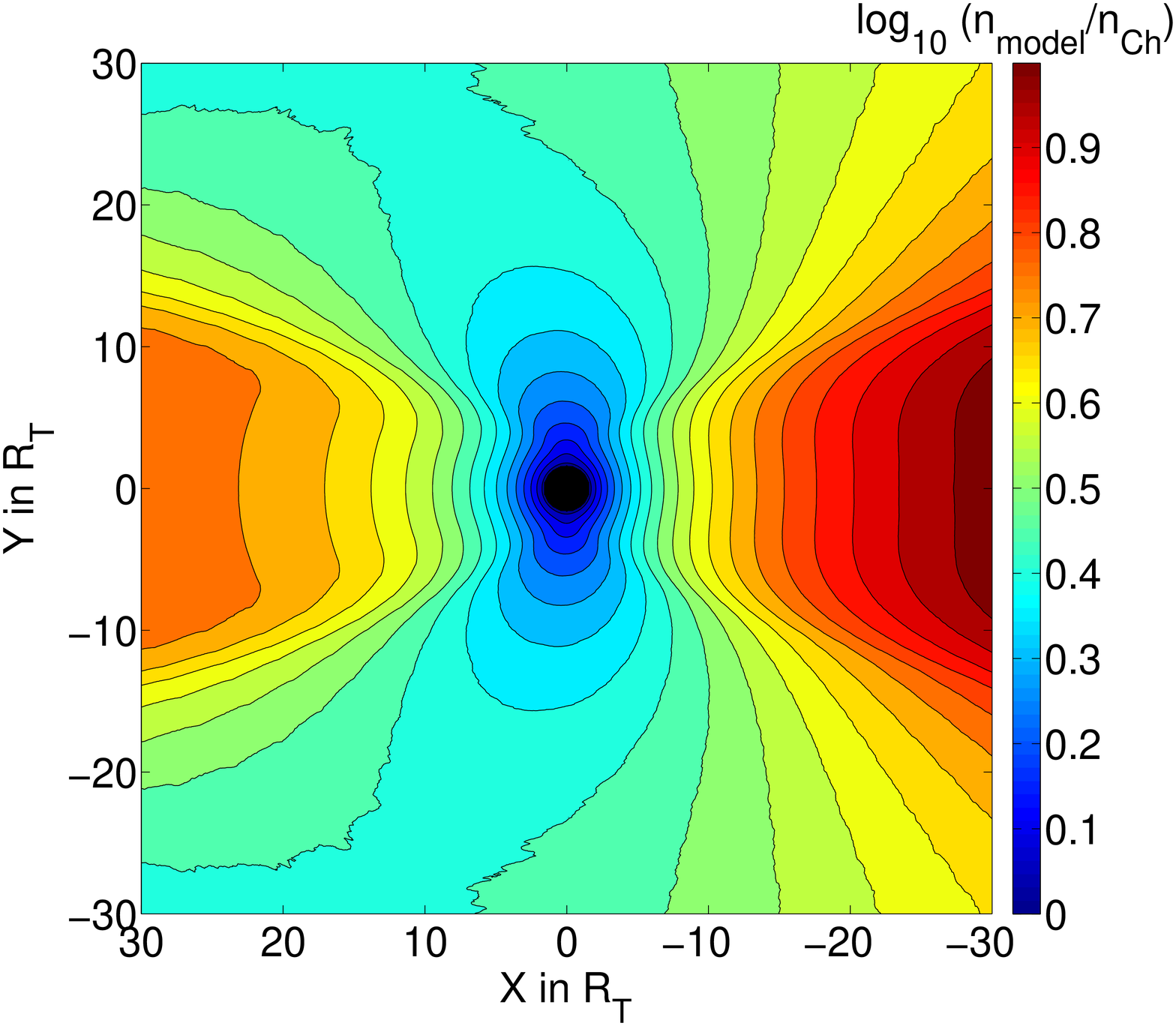}
\caption{Maps of ballistic particles densities at Earth, Mars and Titan (normalized by the exobase density) with an exospheric temperature of respectively $T_{exo}=800\ K$, $200\ K$ and $152\ K$ (from the left to the right). The upper panels represent the output of our model. The middle panels represent with the same colors the ballistic density map provided by the \cite{Chamberlain1963} model. The lower panels represent the ratio between the two models in log-scale: red areas show where our density exceeds the Chamberlain's one, the blue areas represent where the Chamberlain's model overestimates the density.}
\label{figure317}
\end{figure}

First of all, as shown on the figure \ref{figure317} (lower panels), close to the planet, our densities (which include the radiation pressure effect) overcome the ballistic densities from the \citet{Chamberlain1963} model. In the nightside direction at Earth, the densities are up to $10^{0.6}\approx6$ times higher than predicted by \citet{Chamberlain1963}. However, in the dayside direction at Earth (see figure \ref{comparaison}), the densities are less high, with an enhancement factor of about $10^{0.4}\approx 2.5$ times. This is also the case at Titan and Mars (see figure \ref{comparaison}). We will also later see (fig. \ref{prad_faible}) that, even for relative small radiation pressures (i.e. $r_{exo}<10-20\ R_{pressure}$), our density overcomes the one obtained through the \citet{Chamberlain1963} formalism. A physical explanation will be brought in section \ref{satpart}.

In the figure \ref{figure317} that shows the 2D maps of ballistic particles densities at Earth, we see clearly the asymmetries induced by the radiation pressure. As expected and previously mentioned, the ballistic density is clearly more significant in the nightside direction than in the dayside direction. Most of the particles are preferentially blown behind the planet. This observation is in agreement with the work by \cite{Bishop1989}, who also found such an asymmetry corresponding to the well known geotail phenomenon of enhanced nightside densities observed (\citet{Thomas1972}, \citet{Bertaux1973}, \citet{Zoennchen2011}, \citet{Bailey}). Nevertheless, the densities in the dayside direction are not so low: the density profile in the dayside direction remains significant compared with the transverse direction, the Dusk/Dawn/North Pole/South Pole plane. For example, at 10 Earth radii the densities in the dayside direction are about $25\%$ larger than in the perpendicular plane (and $50-100\%$ larger in the nightside direction). The exosphere seems to have a prolate shape oriented along the Sun-planet axis. 

\begin{figure}[!h]
\centering
\includegraphics[width=\linewidth]{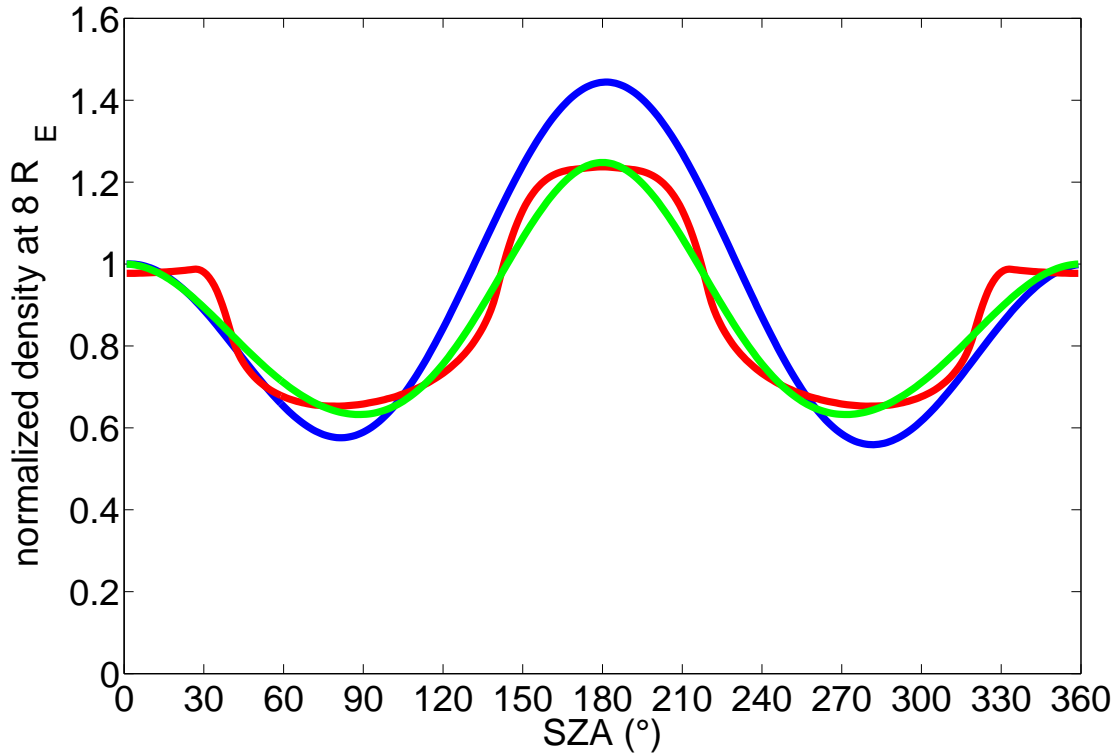}
\caption{Comparison at Earth between our balistic particles density and the observations made by \cite{Zoennchen2013} at the same distance ($8\ R_{E}$), as a function of the Solar Zenithal Angle (SZA) in degrees. We fitted our model (red) with four spherical harmonics (green) as done by \cite{Zoennchen2013} (blue, with only two sinusoids and the constant). For a better comparison, we scaled the blue and green plots in order to be at 1 for SZA$=0$.}
\label{figure319}
\end{figure}

The exospheric density asymmetries were recently observed thanks to the TWINS mission and published by \citeauthor{Zoennchen2011} (\citeyear{Zoennchen2011}, \citeyear{Zoennchen2013}) and \cite{Bailey}. In the figure \ref{figure319}, we compare at constant distance ($8\ R_E$) the variability of the Earth exospheric density from our model and from the observations as a function of the solar zenith angle. As \citeauthor{Zoennchen2011} (\citeyear{Zoennchen2011}, \citeyear{Zoennchen2013}) and \cite{Bailey} fit their observations with a limit number of spherical harmonics, we do the same here for a clear comparison. \citet{Zoennchen2013} observed a decrease of $45\%$ in the dusk and dawn directions in the equatorial plane and an enhancement by up to $45\%$ in the nighside direction compared with the dayside. In the figure \ref{figure319}, the decrease for our model is about $36 \%$ and the increase about  $25\%$. The lower difference in the nightside direction can be explained by the lack of escaping particles in our model, which would certainly increase the density in the nightside direction. 

Our model thus reproduces quite well the local time variation in the equatorial plane although we do not know the precise conditions in the exosphere during the observations by \citet{Zoennchen2013}, i.e. the exact exospheric temperature and radiation pressure acceleration (which is known to have a strong variability (\cite{Vidal1975}). Nevertheless, we have significant discrepancies (compared with \citet{Zoennchen2013}) in the meridional plane in particular during the equinox. One explanation can be the satellite particles as will be discussed in section \ref{satpart}.

\subsection{The exopause}\label{exopause}

\begin{figure}[!h]
\centering
\includegraphics[width=\linewidth]{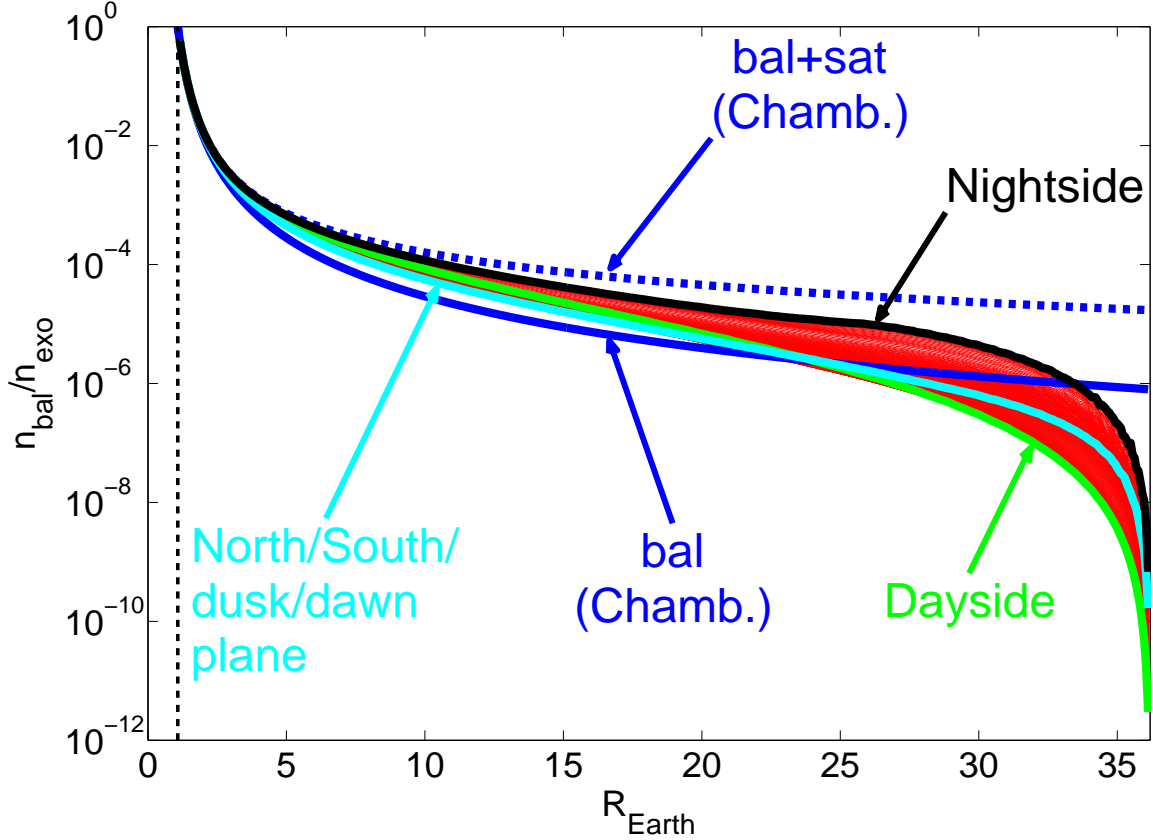}
\caption{Radial density profiles for ballistic particles at Earth with $T_{exo}=800\ K$. The specific directions are in black for the nightside, in green for the dayside and in cyan for the perpendicular plane, whereas a set of profiles for intermediate directions are plotted in red. The corresponding \citeauthor{Chamberlain1963} ballistic and ballistic plus satellite particles densities are given in solid and dashed blue.}
\label{radial}
\end{figure}

We also plotted the ballistic density profiles at Earth as a function of the distance for different $\theta$ (angle with the Sun-planet axis) in the figure \ref{radial}.
Beyond the clear local time asymmetries and enhanced densities compared with the Chamberlain (pure gravity) profiles, we observe a cut-off of the ballistic density at 36 Earth radii. This limit corresponds to an exopause located at $R_{pressure}=\sqrt{{GM}/{a}}=36\ R_{E}$, which was not introduced artificially as was previously done by the previous works (e.g. \citet{Bishop1991}). Unfortunately, we could not demonstrate this property analytically but the modeling proves that the bounded trajectories occur only inside the sphere of radius $R_{pressure}$. With the \citeauthor{Chamberlain1963} approach, the limit for bounded trajectories is the infinity. Physically, a first limit is however imposed by the presence of the Sun (or by nearby planets for satellites) : the Hill's sphere, that defines the limit of the gravitational influence of the central body. But here, the radiation pressure induces another limit (below the Hill sphere radius) located at $R_{pressure}$, i.e. where the accelerations due to the gravity and radiation pressure forces are equal. Beyond this limit, all particles are escaping. In future works, we will show its implication on the evolution of planetary atmospheres.

The sharp density drop (shown for ballistic populations) at the distance $R_{pressure}$ should still be seen after adding the satellite and escaping populations, and lead to a sharp drop in the total density profiles. The densities of satellite and escaping populations are indeed decreasing with distance : the escaping component should approximately decrease in $r^{-2}$, whereas the satellite component will also disappear at Rpressure since no bound motion can exist beyond this limit (see also \citet{Beth2014}). Thus, no strong increase for these two populations will be able to counter the sharp drop of the ballistic component, which should be seen in the total density measured (unless an external source of neutrals, e.g. the neutral part of the solar wind, adds a significant density that hides the sharp drop).

As far as we know, the only relevant neutral density measurement allowing to investigate this topic is the study by \citet{Brandt2012}. These authors provided energetic neutral atom (ENA) images of the Titan environment, and showed that ENA fluxes may be observed up to 50 000 km, i.e. the Hill sphere radius determined by the gravitational influence of Saturn. The ENAs are produced by charge exchange reactions between the magnetospheric ions and the exospheric neutrals, so that the ENA flux profiles provide information about the neutral density profiles. At Titan, the location of the exopause, where a sharp density drop is expected, is due to the gravitational influence of Saturn rather than to the radiation pressure that leads to an $R_{pressure}$ distance much further. The observation of ENA fluxes up to the exopause is thus in agreement with a sharp neutral density drop at the exopause distance expected by our model.

\subsection{The satellite particles}\label{satpart}

The figure \ref{radial} also provides a comparison between our modeled ballistic profiles and the ballistic and ballistic+satellite densities from \cite{Chamberlain1963}. Near the planet, our ballistic density profile remains for any direction between these last two profiles. Why? 

The radiation pressure disturbs the trajectories of particles that are initially conics. As shown by \cite{Bishop1989}, the particles crossing the Sun-planet axis ($P_{\phi}=0$) can not be satellite particles: the bounded trajectories see their periapsis decrease with time and they necessarily cross the exobase if $P_{\phi}=0$. The radiation pressure separates the areas where ballistic and satellite particles can exist: the ballistic particles are preferentially near the Sun-planet axis whereas the satellite particles are mostly located far from this axis in the perpendicular plane (see below). The radiation pressure will convert satellite trajectories into ballistic trajectories and these last ones into escaping particles. This is why we have our density between ballistic and ballistic plus satellite particles densities from \cite{Chamberlain1963}: a part of our ballistic particles were probably initially satellite particles converted into ballistic ones by the radiation pressure. Moreover, such a conversion of trajectories is stronger near the axis than in the perpendicular plane. 
To support this explanation, we cannot calculate the satellite particles density since the Liouville theorem can not be applied to particles which do not cross the exobase. Nevertheless, we can estimate the phase space volume dedicated to the satellite particles, and look where it is maximum. The phase space volume is given by:
\begin{equation*}
V=\int_{\mathbb{R}^3}\!\mathds{1}_{type}\,\mathrm{d}^3\vec{p}
\end{equation*}

\begin{figure}[!h]
\centering
\includegraphics[width=\linewidth]{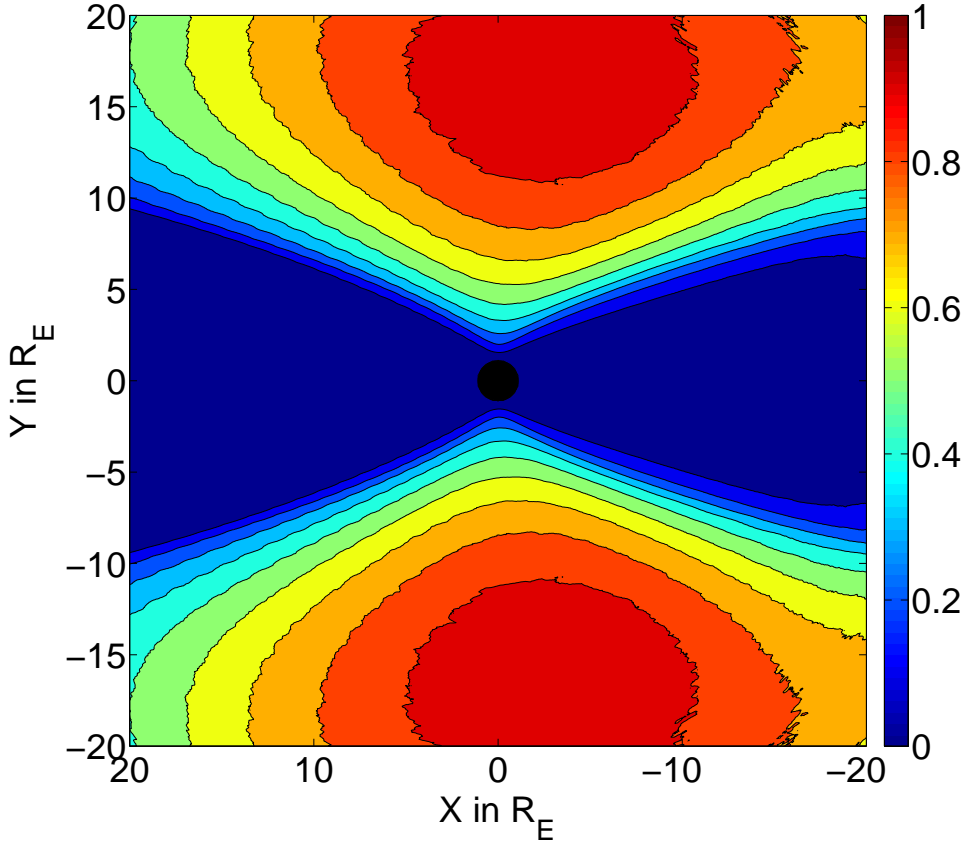}
\caption{Ratio between the phase space volumes for satellite particles and for ballistic particles at Earth with $T_{exo}=800\ K$. In red, the ratio is around 1 and in blue the ratio is around 0. Redder areas are most probable areas where satellite particles can exist.}
\label{volume}
\end{figure}

The bounds of integration (or the value of $\mathds{1}_{type}$) depend on the type of particles considered (type=ballistic or satellite; for escaping particles, $V$ is infinite). This volume has no physical meaning but provides the available volume. However, one can extract a useful information if we compare the phase space volumes for satellite and ballistic particles (given in fig. \ref{volume}). The satellite particles are thus preferentially located far from the Sun-planet axis and in the perpendicular plane.

The existence and production/loss of satellite particles is however not well-known. A recent work by \citet{Beth2014} showed that the satellite particles are produced by scarce collisions just above the exosphere. In the Earth case, they showed the satellites particles do not contribute significantly to exospheric densities. Nevertheless, we mentioned above a discrepancy between our model and the observations by \citet{Zoennchen2013} in the perpendicular plane during equinox, that could be related to the presence of satellite particles. During this period, the polar cusps are in the perpendicular plane. The production of satellite particles might be in the polar cusps where the densities are large and where the satellite trajectories are more stable.

\begin{figure}[!h]
\centering
\includegraphics[width=\linewidth]{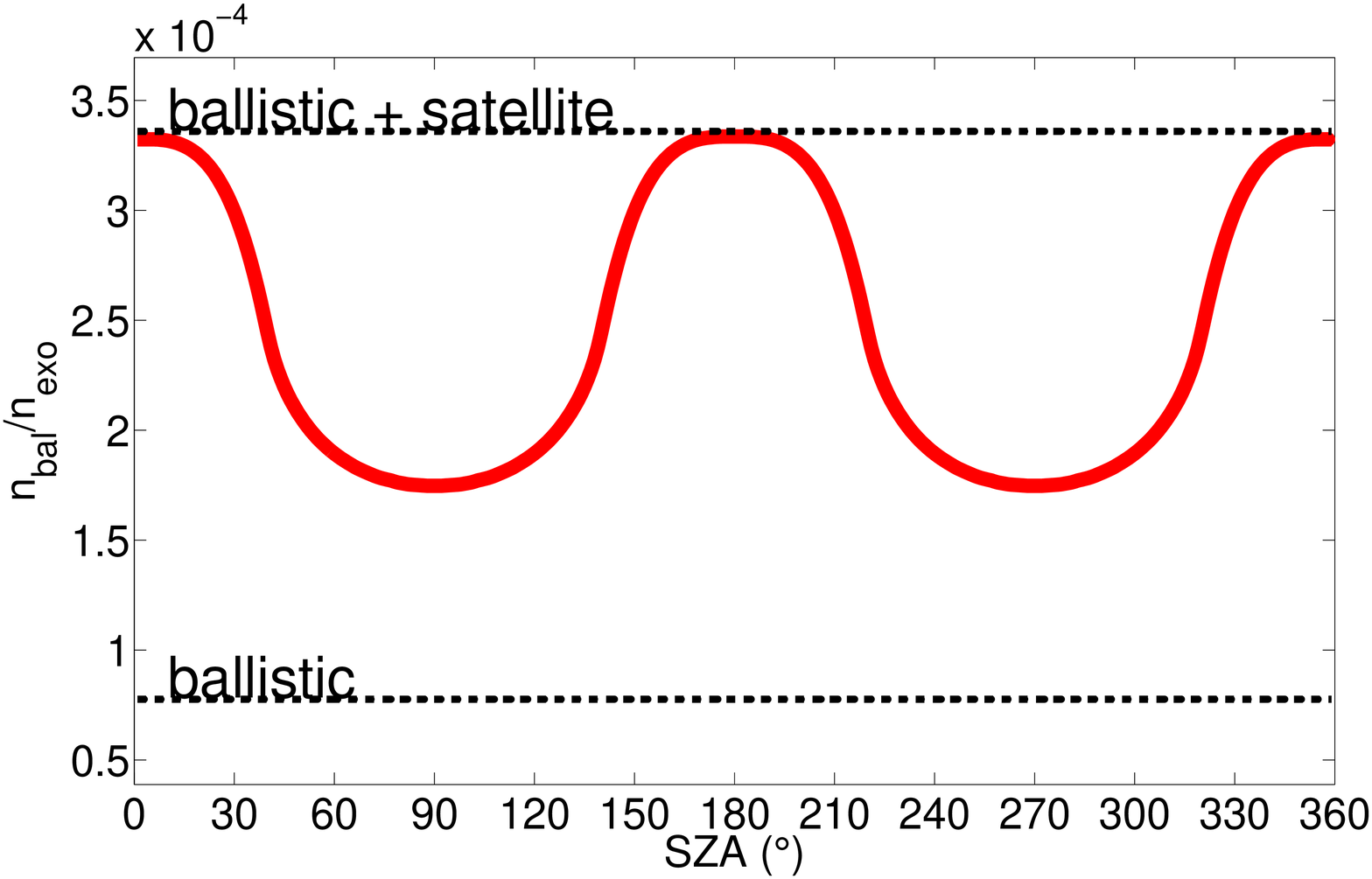}
\caption{Same plot as for the figure \ref{figure319} with a radiation pressure thousand times smaller. We superimposed the densities for ballistic and ballistic plus satellite particles provided by \citet{Chamberlain1963}.}
\label{prad_faible}
\end{figure}

A last evidence for the conversion between ballistic and satellite particles is given by the figure \ref{prad_faible}, where the Earth ballistic density profile at 8 $R_E$ (as in figure \ref{figure319}) is given using a fictive radiation pressure value divided by 1000. We would expect a density profile similar to the pure gravity case, i.e. the Chamberlain profile of ballistic particles. This is however not the case. The radiation pressure force disturbs the trajectories regardless of the value considered: the particles that are quasi-satellite see their periapsis altitude decreases more slowly if the radiation pressure is smaller, so that they can turn around the planet during a longer time without crossing the exobase than with a larger radiation pressure, but they will anyway cross it after some time. In this work, there is no distinction between these quasi-satellite particles and ballistic particles since no simple physical parameter allows to distinguish them.

According to the densities profiles, given in figures \ref{radial} and \ref{prad_faible}, we can reasonably assume that for small distances compared with $R_{pressure}$ (generally some planetary radii except for Hot Jupiters for where the exopause may be located below the exobase), the densities are between $n_{bal}$ and $n_{bal}+n_{sat}$ from the Chamberlain formalism. Even if the collisions are not included in the model (which are the source for satellite particles), assuming Chamberlain density profiles with and without the satellite particles contribution can give a range for exospheric densities at a given distance for the disturbed case by the radiation pressure.

\subsection{Comparative planetary science}

\begin{figure}[!h]
\centering
\includegraphics[width=\linewidth]{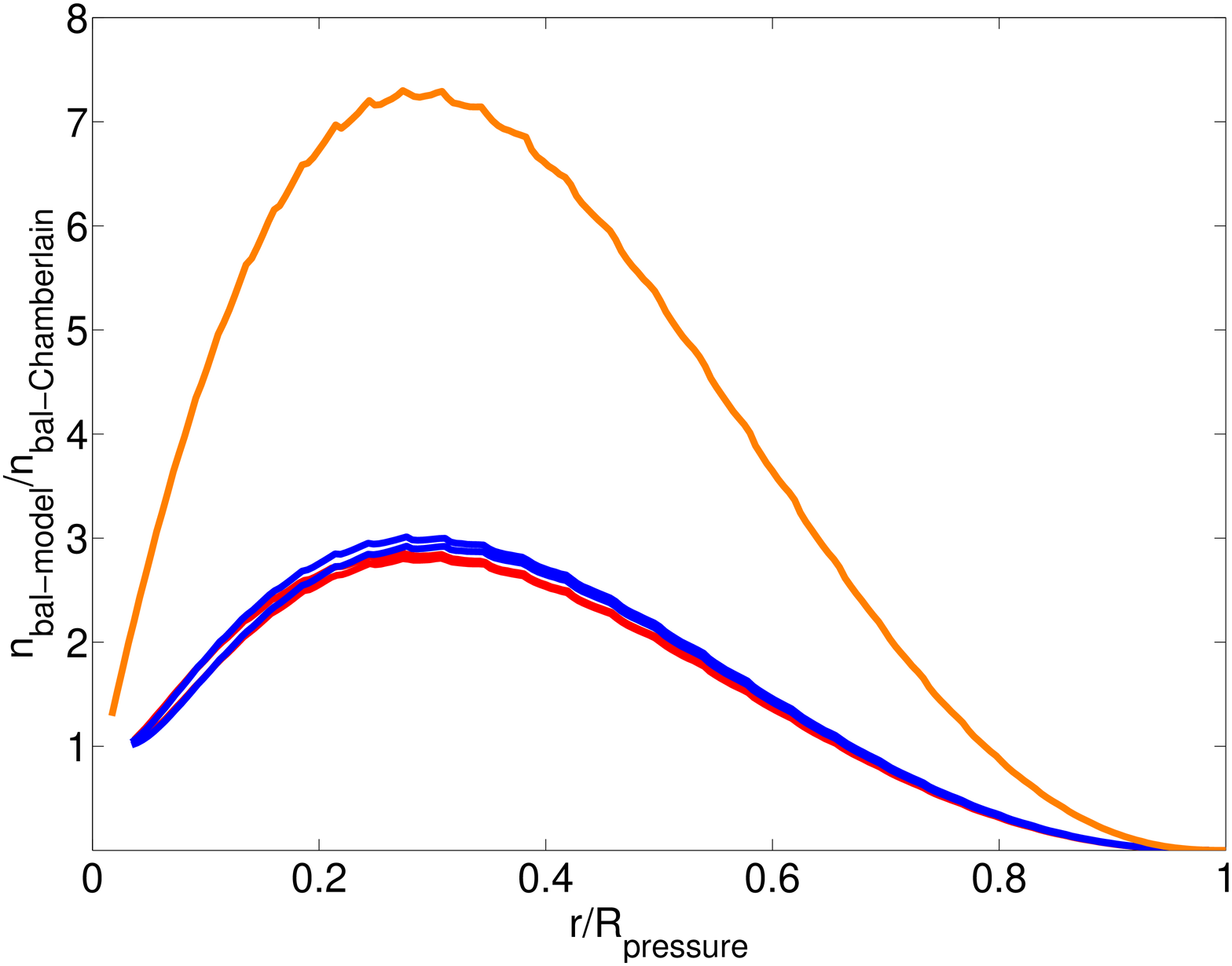}
\caption{Ratio between our ballistic density profile in the dayside direction and the ballistic density profile from \citet{Chamberlain1963} at Earth (blue), Mars (red) and Titan (orange). The two lines for Earth and Mars correspond to two exospheric temperatures: $800\ K$ and $1200\ K$ at Earth, $200\ K$ and $350\ K$ at Mars. For Titan, the exospheric temperature is $152\ K$. The distance $r$ is normalized with the value of the exopause distance (i.e. $R_{pressure}$) for each case.}
\label{comparaison}
\end{figure}

With our studies of different cases, Earth, Mars and Titan, corresponding to different conditions (i.e. exospheric temperatures and radiation pressures), we are able to explore a wide range of configurations and to derive general conclusions.

All our case studies showed, with an exopause (always located at $R_{pressure}$) above the exobase, an enhancement of the ballistic densities compared with those provided by the \citet{Chamberlain1963} model. As shown by the figures \ref{figure317} and \ref{comparaison}, the radiation pressure indeed strongly affects the density profiles. The figure \ref{comparaison} reveals a ratio between the disturbed and non disturbed densities that is very similar for Earth and Mars (up to 3), but larger at Titan (up to 7), with a peak at the same normalized distance. The variability of the induced disturbance between the various planets may be explained as follows. The first effect of the radiation pressure is to break the spherical symmetry: the physical consequence is the conversion of satellite particles into ballistic ones as shown by the figure \ref{prad_faible}. The secondary consequence depends on the intensity of the radiation pressure: the stronger it is (i.e. the closer to the Sun the planet is), the more ballistic particles become escaping ones. We should thus expect less ballistic particles at Earth than at Mars. However, the equations driving the density profiles mostly depend on the parameters $r_{exo}$ and $R_{pressure}$ (i.e. $\lambda_c$ and $\lambda_a$), in particular on the ratio $r_{exo}/R_{pressure}$. Earth and Mars actually have similar ratios $r_{exo}/R_{pressure}$ (i.e. about 35) and thus have a similar enhancement as a function of $r/R_{pressure}$ (cf. figure \ref{comparaison}). For the Titan case, the radiation is hundred times weaker, which would suggest a weak density enhancement, but the ratio $r_{exo}/R_{pressure}$ is completely different: $R_{pressure}$ is $100$ times larger than $r_{exo}$.
Nevertheless, we should take precautions the Titan case because another external force could strongly affect the dynamic of atmospheric species: the gravitational attraction by Saturn. 

\section{Conclusions}\label{conclusions}

In this paper, we generalize the initial work by \citet{Bishop1989} by developing a 2D model (3D if we do not assume the axisymmetry) for the density profiles of ballistic exospheric neutral particles, in order to study the impact of the radiation pressure on the structure of planetary exospheres such as at Earth, Mars or any planet with a dense atmosphere. 

We reproduce quite well with our simulations the different exospheric asymetries observed at Earth:
\begin{itemize}
\item the ``tail phenomenon": the Earth exosphere has higher densities for atomic Hydrogen in the nightside direction than in the dayside direction. This is already known (\citet{Thomas1972},\citet{Bertaux1973}) and directly attributed to the radiation pressure.
\item dusk/dawn/North Pole/South Pole asymmetries (\citet{Bailey2011},\citeauthor{Zoennchen2011}(\citeyear{Zoennchen2011},\citeyear{Zoennchen2013}): the radiation pressure induces a depletion of particles in the perpendicular plane, observed in the equatorial plane.
\end{itemize}

Moreover,  the radiation pressure entails an increase of ballistic particles densities which are in the lower corona (up to several planetary radii), the main exospheric component compared with satellite and escaping particles. Compared with the \citeauthor{Chamberlain1963} model (i.e. without the radiation pressure), the densities are several times higher (up to $2.5$ in the dayside direction and $4$ in the nightside one at Earth). Only the ballistic (i.e. not the escaping) particle density calculation is performed for numerical reasons (time and precision issues).

We highlight also the appearance of a characteristic distance of the exosphere: the ``exopause". This concept was introduced by \citet{Bishop1991}, who included it artificially in their model: this boundary is the limit where the intensity of the radiation pressure is equal to the gravitational attraction. As shown in this paper, this limit appears naturally in our simulations with a break in our density profiles. Physically, the exopause divides the exosphere into two regions: below the exopause, we can find bounded (satellite and ballistic particles) and unbounded (escaping) trajectories; above the exopause, we find only unbounded trajectories. The exopause will lead to a local sharp drop for the total density (including ballistic, satellite and escaping populations), that is in agreement with the observation of energetic neutral atom fluxes at Titan up the Hill sphere radius only (i.e. the exopause for Titan) by \citet{Brandt2012}. We precise that our model provides only density profiles for a planetary exosphere where the exopause is located above the exobase (a future paper will investigate the extreme case of Hot Jupiters where the strong radiation pressure pushes the exopause down to the exobase). The exopause boundary also induces a constraint for modeling the exospheres: the size of the simulations box must be large enough to contain the exopause in order to take into account all the asymmetries induced by this force.

Moreover, in our study, we have also shown the influence of the radiation pressure on the repartition of ballistic and satellite populations. On the one hand, near the Sun-planet axis, the periapsis of the bounded particles decreases slowly because of the radiation pressure until it crosses the exobase. Thus, we find essentially ballistic particles (called also satellite particles with finite lifetime) near the Sun-planet axis. On the other hand, the available regions to find bounded trajectories which do not cross the exobase (i.e. satellite particles) are essentially in the dusk/dawn/North Pole/South Pole plane. Thus, the radiation pressure separates clearly the ballistic and satellite particles regions even if these particles are both bounded. 

Finally, the radiation pressure induces an important effect on the velocity phase space: the radiation pressure converts a part of the satellite particles into ballistic particles and these ones into escaping particles. This explains the enhancement of ballistic densities compared with the \citet{Chamberlain1963} model: the radiation pressure converts efficiently the satellite particles of his model (without radiation pressure) into ballistic ones for ours (with radiation pressure). This explanation is supported by the figures \ref{radial} and \ref{prad_faible}: near the planet or for small radiation pressure accelerations, the densities remain between the ballistic and ballistic plus satellites ones provided by the \citet{Chamberlain1963} formalism. Consequently, including or not the \citeauthor{Chamberlain1963} satellite particles partition function provides an appropriate range of densities to include the radiation pressure effect on exospheric density profiles, even in the absence of collisions that are the source of satellite particles.

In future works, we will study the photogravitationnal Circular Three-Body Problem and the implication on the stability of planetary exospheres, in particular for Hot Jupiters. Moreover, the escaping particles density could not be calculated here because of numerical issues but the escaping flux will be investigated in details in a future work.

\bibliographystyle{model2-names}

\end{document}